\documentclass[11pt,a4paper]{article}
\pdfoutput=1
\usepackage{jheppub}

\usepackage{verbatim}
\usepackage{listings}
\usepackage{fancyvrb}
\usepackage{xcolor}
\usepackage{fancyvrb}

\usepackage[T1]{fontenc}
\usepackage{graphicx}
\usepackage{slashed}
\usepackage{amsmath,amsfonts,bbm,latexsym,amssymb}
\usepackage{epsfig}
\usepackage{array}

\usepackage{booktabs}
\usepackage[utf8]{inputenc}
\usepackage{fancyvrb}
\usepackage{subfig}

\usepackage{graphicx}
\usepackage{caption}
\usepackage{feynmp-auto}

\usepackage[flushleft]{threeparttable}
\makeatletter
\def\hlinewd#1{%
	\noalign{\ifnum0=`}\fi\hrule \@height #1 %
	\futurelet\reserved@a\@xhline}
\makeatother
\usepackage{multirow}
\usepackage{multicol}
\usepackage{longtable}

\usepackage[b]{esvect}
 
\usepackage{hyperref}
\hypersetup{colorlinks=false}
\usepackage{array}

\usepackage{titlesec}
\usepackage{dashrule}
\usepackage{scrextend}
\usepackage{ragged2e}
\usepackage{cancel}
\usepackage{float,hypcap}
\usepackage{changepage}
\usepackage{sectsty}
\usepackage{soul}
\usepackage{bbold}
\usepackage{slashed}
\usepackage{tabularx}
\usepackage{blkarray}
\usepackage{latexsym}
\usepackage{braket}
\usepackage{enumerate}
\usepackage{bm}
\usepackage{booktabs,hyperref}
\usepackage{bigints}
\usepackage{amsmath}
\usepackage{amssymb}
\usepackage{empheq}
\usepackage{stackengine}
\stackMath

\usepackage{color}
\definecolor{orange}{rgb}{1,0.7294,0.1843}
\definecolor{uglyblue}{RGB}{95,158,160}
\definecolor{mygray}{RGB}{129,129,129}
\definecolor{mywhite}{RGB}{255,250,240}
\usepackage[most]{tcolorbox}

\allowdisplaybreaks

\def\be{\begin{equation}}
\def\ee{\end{equation}}
\def\ba{\begin{alignedat}}
\def\ea{\end{alignedat}}
\def\bea{\begin{eqnarray}}
\def\eea{\end{eqnarray}}
\newcommand{\bs}{\begin{subequations}}
\newcommand{\es}{\end{subequations}}

\def\hs{\hspace}

\def\no{\nonumber\\}
\def\fn{\footnote}

\def\t{\texttt}
\def\ts{\textsc}

\def\FM{\textsc{FeynMaster}}
\def\FMS{\textsc{FeynMaster} }

\newcommand{\newc}{\newcommand}
\newc{\ol}{\overline}
\newc{\wt}{\widetilde}

\newc{\m}{\mathcal}

\newcommand{\beq}{\begin{eqnarray}}
\newcommand{\eeq}{\end{eqnarray}}
\newcommand{\bpmatrix}{\begin{pmatrix}}
\newcommand{\epmatrix}{\end{pmatrix}}

\renewcommand{\ol}{\text{1l}}

\renewcommand{\eqref}[1]{eq.~(\ref{#1})}

\newcommand{\bc}{\begin{center}}
\newcommand{\ec}{\end{center}}

\newcommand{\gsim}{\raisebox{-0.13cm}{~\shortstack{$>$ \\[-0.07cm]
      $\sim$}}~}

\def\m#1{m_{#1}}

\def\ltap{\;\centeron{\raise.35ex\hbox{$<$}}{\lower.65ex\hbox{$\sim$}}\;}
\def\gtap{\;\centeron{\raise.35ex\hbox{$>$}}{\lower.65ex\hbox{$\sim$}}\;}

\usepackage{stackengine}
\stackMath

\begin{document}

\begin{flushright}
CFTP/22-001
\end{flushright}

\vspace{-5mm}

\title{\boldmath The one-loop impact of a dependent mass: the role of $m_3$ in the C2HDM}

\author[a,b]{Duarte Fontes,}
\author[b]{Jorge C. Rom\~ao}
\affiliation[a]{Department of Physics, Brookhaven National Laboratory, Upton, NY, 11973, USA}
\affiliation[b]{Departamento de F\'{\i}sica and CFTP,
Instituto Superior T\'{e}cnico, Universidade de Lisboa, \\
Avenida Rovisco Pais 1, 1049-001 Lisboa, Portugal}
\emailAdd{dfontes@bnl.gov}
\emailAdd{jorge.romao@tecnico.ulisboa.pt}

\vspace{0.5cm}
\noindent\abstract{
In the complex 2-Higgs-Doublet Model (C2HDM), the mass $m_3$ of the heaviest neutral scalar $h_3$ is usually chosen as a derived parameter. We investigate one-loop corrections to $m_3$ and their impact on decays of $h_3$. 
Very fine-tuned regions of the parameter space can be found where such corrections are large, not due to subtraction schemes, but rather due to the particular dependence of $m_3$ on the independent parameters.
We show that even moderate corrections can have a significant impact on decays of $h_3$, as they may be several times enhanced by leading-order factors. 
}

\maketitle
\noindent

\section{Introduction}

After the observation of the Higgs boson at the Large Hadron Collider (LHC) in 2012~\cite{Aad:2012tfa, Chatrchyan:2012ufa}, it is of the utmost importance to further explore the scalar sector of particle physics. Particularly relevant is the possibility that such sector is extended when compared to that of the Standard Model (SM); examples of models with an extended scalar sector can be found e.g. in refs. \cite{Gunion:1989we,Ivanov:2017dad,Branco:2011iw}. One of the most common models---the so-called real 2-Higgs-Doublet model (2HDM), where CP is assumed to be conserved in the potential---was recently suspected of theoretical unsoundness  \cite{Fontes:2021znm},
as the simultaneous enforcement of CP conservation in one sector of the model (the potential) and the allowance of CP violation in another one (quark interactions) will likely end up leading to divergent predictions.
The model that heals it in the simplest fashion---while accounting for the already observed CP violation---is the complex 2HDM (C2HDM)
\cite{
Ginzburg:2002wt,
Khater:2003wq,
ElKaffas:2006gdt,
WahabElKaffas:2007xd,
ElKaffas:2007rq,
Osland:2008aw,
Grzadkowski:2009iz,
Arhrib:2010ju,
Barroso:2012wz,
Inoue:2014nva,
Cheung:2014oaa,
Fontes:2014xva,
Chen:2015gaa,
Fontes:2015xva,
Fontes:2017zfn,
Boto:2020wyf,
Cheung:2020ugr,
Fontes:2021iue,
Fontes:2021kue,
Frank:2021pkc},
where CP is explicitly violated in the scalar sector at tree-level (thus providing a new source of CP violation, as required by the three Sakharov criteria for baryogenesis \cite{Sakharov:1967dj}).
Accordingly, the three neutral scalars described by the C2HDM ($h_1$, $h_2$, $h_3$) contain in general a mixture of a CP-odd component and a CP-even one. LHC searches have already excluded the possibility according to which the scalar boson discovered in 2012 is a pure CP-odd state \cite{Chatrchyan:2012jja,Khachatryan:2014kca,Aad:2013xqa}; yet, the scenario where it contains a CP-odd fraction is still allowed \cite{Zyla:2020zbs}.

As the LHC run-3 approaches, deviations from SM predictions are anxiously expected. The absence of any undoubtable signal of new physics so far implies that such deviations shall be subtle. As a consequence, precise predictions from the theory side are necessary. These require the inclusion of one-loop electroweak corrections, which in turn require the one-loop electroweak renormalization of the model at stake. We have recently presented the one-loop renormalization of the C2HDM \cite{Fontes:2021iue}, and showed that it configures a rather unique program of renormalization. Indeed, there are more independent counterterms than independent renormalized parameters, which means that several combinations of the former can be chosen for the same set of the latter. Moreover, the parameter $m_3$ (corresponding to the mass of the heaviest neutral Higgs boson, $h_3$) is taken as a dependent parameter, so that it cannot be fixed to be equal to the physical mass.

In this paper, we investigate the one-loop corrections to $m_3$ in a particular variant of the C2HDM, the Type II C2HDM. Higher order corrections to dependent masses have been discussed at length in the context of other models, such as supersymmetric extensions of the SM (cf. ref. \cite{Slavich:2020zjv} and references therein). The importance of the investigation of the higher order corrections to the mass of $h_3$ in the C2HDM is at least twofold: first, it allows for a refined search of $h_3$, whose physical mass may contain a non-negligible one-loop contribution; secondly, it opens the possibility of calculating one-loop corrections to processes with an external $h_3$. In fact, although the residue of $h_3$ can be set to unity (implying that the wave-function renormalization factors become trivial), there will be non-trivial contributions to the decay width of processes with an external $h_3$ due to the corrections to $m_3$. We explicitly address the last point by considering the next-to-leading order decay width of
several $h_3$ decays.

The paper is organized as follows: we begin by presenting a theoretical setup of the C2HDM in section \ref{sec:setup}, especially focused on the mass and decays of $h_3$; then, after discussing in section \ref{sec:procedure} the software and the simulation procedure used, we present our results in section \ref{sec:results}. We finish the paper with some conclusions in section \ref{sec:conclusions}.



\section{Theoretical setup}
\label{sec:setup}

\subsection{The model at tree-level}
\label{sec:model}

Details on the Type II C2HDM  and its renormalization can be found in refs. \cite{Fontes:2021iue,Fontes:2021kue}. Here, we summarize the most relevant aspects for the analysis of the mass and decays of $h_3$.
We write the potential as:
\bea
\label{eq:mypot}
V &=& m_{11}^2 |\Phi_1|^2 + m_{22}^2 |\Phi_2|^2
- \left(m_{12}^2 \, \Phi_1^\dagger \Phi_2 + \text{h.c.}\right)
+ \frac{\lambda_1}{2} (\Phi_1^\dagger \Phi_1)^2 +
\frac{\lambda_2}{2} (\Phi_2^\dagger \Phi_2)^2 \nonumber \\
&& + \lambda_3
(\Phi_1^\dagger \Phi_1) (\Phi_2^\dagger \Phi_2) + \lambda_4
(\Phi_1^\dagger \Phi_2) (\Phi_2^\dagger \Phi_1) +
\left[\frac{\lambda_5}{2} (\Phi_1^\dagger \Phi_2)^2 + \text{h.c.}\right] \;.
\eea
All parameters are forced to be real by hermiticity, except $m_{12}^2$ and $\lambda_5$, which are in general complex. $\Phi_1$ and $\Phi_2$ are the scalar doublets, which we parameterize as:
\be
\Phi_1 =
\left(
\begin{array}{c}
\phi_1^+\\
\tfrac{1}{\sqrt{2}} (v_1 \, e^{i \zeta_1}+ \rho_1 + i \eta_1)
\end{array}
\right),
\hspace{5ex}
\Phi_2 =
\left(
\begin{array}{c}
\phi_2^+\\
\tfrac{1}{\sqrt{2}} (v_2 \, e^{i \zeta_2}+ \rho_2 + i \eta_2)
\end{array}
\right),
\label{eq:doublets-first}
\ee
where $v_i$ are the real vacuum expectation values (vevs), $\zeta_i$ (real) phases, $\rho_i$ and $\eta_i$ real fields and $\phi_i^+$ complex fields ($i=\{1,2\}$).
We introduce the total vev $v$ and the angle $\beta$ obeying the relations:
\be
v^2 \equiv v_1^2 + v_2^2,
\qquad
\tan(\beta) \equiv \dfrac{v_2}{v_1}.
\ee
Since there is CP violation at tree-level in the scalar sector, the fields $\rho_i$ and $\eta_i$ in eq. \ref{eq:doublets-first} will all mix at loop-level. To account for such mixing, one must consider the most general diagonalization at tree-level.
We thus introduce the orthogonal matrix $Q$, such that:
\be
S_n = Q \, \phi_n
\quad
\Leftrightarrow
\quad
\begin{pmatrix}
h_1\\
h_2\\
h_3\\
G_0
\end{pmatrix}
= Q
\begin{pmatrix}
\rho_1\\
\rho_2\\
\eta_1\\
\eta_2
\end{pmatrix},
\label{eq:newmain}
\ee
where $G_0$ is the would-be Goldstone boson, and $h_1$, $h_2$ and $h_3$ are the physical neutral scalar bosons of the model, with masses $m_1$, $m_2$ and $m_3$, respectively, complying to $m_1 < m_2 < m_3$.
The matrix $Q$ is parameterized by six angles---$\alpha_0, \alpha_1, \alpha_2, \alpha_3, \alpha_4, \alpha_5$---, according to:
\be
Q
=
Q_5 \, Q_4 \, Q_3 \, Q_2 \, Q_1 \, Q_0,
\label{eq:Q1}
\ee
with
\bea
&&
Q_5
=
\begin{pmatrix}
1 & 0 & 0 & 0\\
0 & c_{5} & 0 & s_{5}\\
0 & 0 & 1 & 0\\
0 & -s_{5} & 0 & c_{5}
\end{pmatrix},
\ \
Q_4
=
\begin{pmatrix}
c_{4} & 0 & 0 & s_{4}\\
0 & 1 & 0 & 0\\
0 & 0 & 1 & 0\\
-s_{4} & 0 & 0 & c_{4}
\end{pmatrix}
\ \
Q_3
=
\begin{pmatrix}
1 & 0 & 0 & 0\\
0 & c_{3} & s_{3} & 0\\
0& -s_{3} & c_{3} & 0 \\
0 & 0 & 0 & 1
\end{pmatrix},
\nonumber \\[3mm]
&&
Q_2
=
\begin{pmatrix}
 c_{2} & 0 & s_{2} & 0\\
0 & 1 & 0 & 0\\
-s_{2} & 0 & c_{2} & 0 \\
0 & 0 & 0 & 1
\end{pmatrix},
\ \
Q_1
=
\begin{pmatrix}
c_{1} & s_{1} & 0 & 0\\
-s_{1} & c_{1} & 0 & 0\\
0 & 0 & 1 & 0\\
0 & 0 & 0 & 1
\end{pmatrix}
\ \
Q_0
=
\begin{pmatrix}
1 & 0 & 0 & 0\\
0 & 1 & 0 & 0\\
0 & 0 & -s_{0} & c_{0}\\
0 & 0& c_{0} & s_{0}
\end{pmatrix},
\label{eq:Q2}
\eea
with $s_i = \sin \alpha_i$, $c_i = \cos \alpha_i$ ($i = \{0,1,2,3,4,5\}$).
The charged scalars in the mass basis, $G^+$ and $H^+$ (with $G^+$ being the charged would-be Goldstone boson), can be defined in a similar fashion.
It turns out that, among the parameters just introduced, several of them end up being dependent, in such a way that several combinations of dependent parameters can be chosen. Following ref. \cite{Fontes:2021iue}, we focus on the four combinations displayed in table \ref{tab:combos}.
In each combination $C_i$ ($i=\{1,2,3,4\}$), $m_3^2$ is a dependent parameter, and is hence fixed as a function of the independent parameters of the combination at stake.%
\fn{\label{note:cumbersome}The expressions for $m_3^2$ in the different combinations are very cumbersome, so that we omit them here.}
\begin{table}[h!]
\centering
\begin{tabular}
{
@{\hspace{-0.8mm}}
>{\centering}p{3cm}
>{\centering}p{0.5cm}
>{\centering}p{0.5cm}
>{\centering}p{0.5cm}
>{\centering\arraybackslash}p{0.5cm}@{\hspace{3mm}}
}
\hlinewd{1.1pt}
Combination & \multicolumn{4}{c}{Dependent parameters}\\
\hline\\[-3.5mm]
$C_1$ & $m_3^2,$ & $\zeta_1,$ & $\alpha_0,$ & $\alpha_4$\\[1mm]
$C_2$ & $m_3^2,$ & $\zeta_1,$ & $\alpha_0,$ & $\alpha_5$\\[1mm]
$C_3$ & $m_3^2,$ & $\zeta_1,$ & $\alpha_4,$ & $\alpha_5$\\[1mm]
$C_4$ & $m_3^2,$ & $\alpha_0,$ & $\alpha_4,$ & $\alpha_5$\\[0.5mm]
\hlinewd{1.1pt}
\end{tabular}
\vspace{-1mm}
\caption{Dependent parameters for each combination $C_i$.}
\label{tab:combos}
\end{table}
\normalsize

\subsection{The mass of $h_3$}
\label{sec:the-mass-ini}

When the C2HDM is considered not simply at tree-level---or at leading order (LO)---, but 
up to one-loop level---or at next-to-leading order (NLO)---, the original quantities of the model (parameters and fields) are taken to be bare quantities. These are shown in the following with index ``$(0)$'' and are separated into renormalized quantities and counterterms; for example, for the squared mass of $h_3$, 
\be
m_{3(0)}^2 = m_{3\mathrm{R}}^2 + \delta m_{3}^2.
\label{eq:mass-scalar-expa-2}
\ee
In this relation, and contrary to what we do for most of the remaining parameters, the renormalized squared mass $m_{3\mathrm{R}}^2$ explicitly includes an index R. The reason is that, as we saw, the squared mass of $h_3$ was chosen as a dependent parameter, which means that $m_{3(0)}^2$ is also dependent, which in turn implies that both $m_{3\mathrm{R}}^2$ and $\delta m_{3}^2$ are also dependent, i.e. fixed.%
\fn{Finite parts can in principle be freely chosen; however, we will want $m_{3\mathrm{R}}^2$ to correspond to the expression for the mass of $h_3$ which is obtained when the C2HDM is studied solely at tree-level (i.e. without aiming at the renomalization of the theory) \cite{Fontes:2017zfn}, so that we exclude such freedom.}
As a consequence, $m_{3\mathrm{R}}^2$ cannot be set equal to the squared pole mass of $h_3$, which we identify with $m_{3\mathrm{P}}^2$. Up to one-loop level, the two masses are related by \cite{Fontes:2021iue}:%
\fn{We use the definition of real pole mass, also known as Breit-Wigner mass. The operator $\widetilde{\operatorname{Re}}$, which is commonly used in the on-shell subtraction scheme, neglects the imaginary parts of loop integrals, while keeping the imaginary parts of complex parameters.}
\be
m_{3\mathrm{P}}^2 = m_{\mathrm{3R}}^2 - \widetilde{\operatorname{Re}} \, \Sigma^{h_3h_3}(m_{3\mathrm{R}}^2) + \delta m_3^2,
\label{eq:m3P}
\ee
where $\Sigma^{h_3h_3}$ is the non-renormalized one-loop diagonal 2-point function of $h_3$. Now, since the tree-level mass $m_3$ depended on the combination $C_i$ (so that it was given by different expressions in the different combinations), the counterterm $\delta m_3^2$ will also correspond to different expressions according to the combination. More specifically, we can write%
\fn{The notation $\stackrel{C_i}{=}$ means that the equality at stake is only valid for the combination $C_i$. Just as the expressions for $m_3^2$ at tree-level (recall note \ref{note:cumbersome}), the expression for $\delta m_3^2$ in the different combinations are extremely cumbersome; we omit them as well.}
\bs
\label{eq:dm3-combos}
\bea
\delta m_3^2 &\stackrel{C_1}{=}& \delta m_3^2 \, (\delta m_1^2, \delta m_2^2, \delta \beta, \delta \alpha_1, \delta \alpha_2, \delta \alpha_3, \delta \alpha_5), \\
\delta m_3^2 &\stackrel{C_2}{=}& \delta m_3^2 \, (\delta m_1^2, \delta m_2^2, \delta \beta, \delta \alpha_1, \delta \alpha_2, \delta \alpha_3, \delta \alpha_4), \\
\delta m_3^2 &\stackrel{C_3}{=}& \delta m_3^2 \, (\delta m_1^2, \delta m_2^2, \delta \beta, \delta \alpha_1, \delta \alpha_2, \delta \alpha_3, \delta \alpha_0), \\
\delta m_3^2 &\stackrel{C_4}{=}& \delta m_3^2 \, (\delta m_1^2, \delta m_2^2, \delta \beta, \delta \alpha_1, \delta \alpha_2, \delta \alpha_3, \delta \zeta_1).
\eea
\es
%
%
The mass counterterms $\delta m_1^2$ and $\delta m_2^2$ are calculated in the on-shell subtraction (OSS) scheme. In the combination $C_4$ (where $\delta \zeta_1$ is independent), $\delta \zeta_1$ is calculated in the modified minimal subtraction ($\overline{\text{MS}}$) scheme; this will introduce an explicit dependence on the renormalization scale, $\mu_{\mathrm{R}}$. 
We fix the mixing parameters using symmetry relations \cite{Fontes:2021iue}; this requires the introduction of the field counterterms for the scalar fields, according to:
\bs
\label{eq:field-scalar-expa}
\bea
\label{eq:charged-diag-bare-exp}
\begin{pmatrix}
G_{(0)}^{+} \\ H_{(0)}^{+}
\end{pmatrix}
&=&
\begin{pmatrix}
1 + \dfrac{1}{2} \delta Z_{G^{+}G^{+}} & \dfrac{1}{2} \delta Z_{G^{+}H^{+}} \\
\dfrac{1}{2} \delta Z_{H^{+}G^{+}} & 1 + \dfrac{1}{2} \delta Z_{H^{+}H^{+}}
\end{pmatrix}
\begin{pmatrix}
G^{+} \\ H^{+}
\end{pmatrix},
\\
\begin{pmatrix}
h_{1(0)} \\ h_{2(0)} \\ h_{3(0)} \\ G_{0(0)}
\end{pmatrix}
&=&
\begin{pmatrix}
1 + \frac{1}{2} \delta Z_{h_1h_1} & \frac{1}{2} \delta Z_{h_1h_2} & \frac{1}{2} \delta Z_{h_1h_3} & \frac{1}{2} \delta Z_{h_1G_0}\\
\frac{1}{2} \delta Z_{h_2h_1} & 1 + \frac{1}{2} \delta Z_{h_2h_2} & \frac{1}{2} \delta Z_{h_2h_3} & \frac{1}{2} \delta Z_{h_2G_0}\\
\frac{1}{2} \delta Z_{h_3h_1} & \frac{1}{2} \delta Z_{h_3h_2} & 1 + \frac{1}{2} \delta Z_{h_3h_3}  & \frac{1}{2} \delta Z_{h_3G_0}\\
\frac{1}{2} \delta Z_{G_0h_1} & \frac{1}{2} \delta Z_{G_0h_2} & \frac{1}{2} \delta Z_{G_0h_3} & 1 + \frac{1}{2} \delta Z_{G_0G_0}
\end{pmatrix}
\begin{pmatrix}
h_{1} \\ h_{2} \\ h_{3} \\ G_{0}
\end{pmatrix}.
\label{eq:neutral-field-CTs}
\eea
\es
All the field counterterms involved in these equations are fixed in the OSS scheme.%
\fn{As in ref. \cite{Fontes:2021iue}, we extend the original OSS scheme (which applies to physical particles only) to apply also to the would-be Goldstone bosons.}
Then, the relevant counterterms for the mixing parameters are defined as:
\bs
\label{eq:mix-params}
\begin{flalign}
& \delta \beta
= \dfrac{1}{4}
\operatorname{Re}
\left.\Big[\delta Z_{G^+H^+} - \delta Z_{H^+G^+}\Big]
\right|_{\xi=1},
\\[3mm]
&
\label{eq:indep-dalfa0}
\delta \alpha_0
\stackrel{C_3}{=}
\dfrac{1}{4} \sec(\alpha_2) \sec(\alpha_3)  \big( \delta Z_{G_0h_3} - \delta Z_{h_3G_0} \big)
\big|_{\xi=1},
\\[3mm]
&
\delta \alpha_1
=
\dfrac{1}{4} \sec(\alpha_2) \left. \Big[  \cos(\alpha_3) \left( \delta Z_{h_1h_2} - \delta Z_{h_2h_1} \right)  +  \sin(\alpha_3) \left( \delta Z_{h_3h_1} -\delta Z_{h_1h_3} \right)  \Big] \right|_{\xi=1},
\\[3mm]
&
\delta \alpha_2
= 
\left.
\dfrac{1}{4}  \sin(\alpha_3) \Big[ \delta Z_{h_1h_2} - \delta Z_{h_2h_1} + \cot(\alpha_3) \left( \delta Z_{h_1h_3} - \delta Z_{h_3h_1} \right)  \Big] \right|_{\xi=1},
\\[3mm]
&
\delta \alpha_3
=
\dfrac{1}{4} \bigg[\delta Z_{h_2h_3} - \delta Z_{h_3h_2} -   \cos(\alpha_3) \tan(\alpha_2)  \left( \delta Z_{h_1h_2} - \delta Z_{h_2h_1} \right) \nonumber\\[-3mm]
& \hs{50mm} + \sin(\alpha_3) \tan(\alpha_2) \left( \delta Z_{h_1h_3} - \delta Z_{h_3h_1} \right)  \bigg] \bigg|_{\xi=1},
\\[3mm]
&
\label{eq:indep-dalfa4}
\delta \alpha_4
\stackrel{C_2}{=}
\dfrac{1}{4} \bigg[\delta Z_{h_1G_0} -\delta Z_{G_0h_1} + \sec(\alpha_3) \tan(\alpha_2) \left( \delta Z_{G_0h_3} - \delta Z_{h_3G_0} \right)  \bigg]
\bigg|_{\xi=1},
\\[3mm]
&
\delta \alpha_5
\stackrel{C_1}{=}
\dfrac{1}{4} \bigg[\delta Z_{h_2G_0} -\delta Z_{G_0h_2} + \tan(\alpha_3) \left( \delta Z_{G_0h_3} - \delta Z_{h_3G_0} \right)
\bigg]
\bigg|_{\xi=1},
\end{flalign}
\es
where $|_{\xi=1}$ means that the calculation is performed in the Feynman gauge.%
\fn{These counterterms are \textit{defined} in this gauge. Hence, even if the calculation of the $S$-matrix is performed in a different gauge (which implies that the remaining counterterms and the non-renormalized functions are calculated in that gauge), the counterterms in eqs. \ref{eq:mix-params} are nonetheless calculated in the Feynman gauge. In the end, the $S$-matrix elements are ensured to be gauge independent. For details, see ref. \cite{Fontes:2021iue}.}
In this way, and given eqs. \ref{eq:dm3-combos}, it is clear that $\delta m_3^2$ depends on the combination $C_i$. 

Such dependence on the combinations is \textit{not} verified for $m_{\mathrm{3R}}^2$, as we now clarify. The different combinations were introduced because we considered a multiplicity of parameters at tree-level (as in eq. \ref{eq:Q2}). This multiplicity was needed in order to generate all the counteterms required to absorb the one-loop divergences (as in eq. \ref{eq:mix-params}). Now, once the counterterms are generated---by separating the original parameters into renormalized ones and counterterms---, the renormalized parameters are generated as well, and in the same quantity as the counterterms. However, several renormalized parameters can be absorbed away; in fact, since their bare versions were considered with the single purpose of yielding counterterms---and in such a way that, were it not for the counterterms, a small set could have been introduced---, a simpler parameterization can be used for the renormalized parameters.%
\fn{More details can be found in ref. \cite{Fontes:2021iue}. Following this reference, we write the renormalized phase $\zeta_{1\mathrm{R}}$ with an explicit R subscript (see also ref. \cite{Fontes:2021kue}).}
More specifically, the renormalized parameters obey:
\be
\zeta_{1\mathrm{R}} =
\alpha_4 =
\alpha_5 =
0,
\qquad
\alpha_0 = \beta.
\label{eq:key}
\ee
This means, finally, that the renormalized parameters do not depend on combinations. In particular, whichever the combination $C_i$ chosen, the renormalized squared mass of $h_3$ reads:%
\fn{In the so-called real 2HDM, there is no such dependence relation between the masses and the mixing angles, so that all of them can be taken as independent parameters.}
\be
m_{3\mathrm{R}}^2 = \frac{m_1^2\, R_{13} (R_{12} \tan{\beta} - R_{11})
+ m_2^2\ R_{23} (R_{22} \tan{\beta} - R_{21})}{R_{33} (R_{31} - R_{32} \tan{\beta})},
\label{eq:m3_derived}
\ee
where the matrix $R$ is given by:
\be
R =
\left(
\begin{array}{ccc}
c_1 c_2 & s_1 c_2 & s_2\\
-(c_1 s_2 s_3 + s_1 c_3) & c_1 c_3 - s_1 s_2 s_3  & c_2 s_3\\
- c_1 s_2 c_3 + s_1 s_3 & -(c_1 s_3 + s_1 s_2 c_3) & c_2 c_3
\end{array}
\right).
\label{eq:matrixR}
\ee
We will be interested in studying the relative corrections to the mass of $h_3$, defined as:
\be
\Delta m_3 \equiv \dfrac{m_{3\mathrm{P}} - m_{3\mathrm{R}}}{m_{3\mathrm{R}}}.
\label{eq:Deltam3}
\ee
Note that this quantity depends on the combination $C_i$, since $m_{3\mathrm{P}}$ depends on $C_i$ (through $\delta m_3^2$, recall eq. \ref{eq:m3P}).

\subsection{NLO decay widths of $h_3$}
\label{section:decays-expressions}

We now focus on decays of $h_3$. 
If $j$ represents the process corresponding to a certain decay of $h_3$, the renormalized NLO amplitude for $j$ can be written as:
\be
\hat{\mathcal{M}}_{j} = \mathcal{M}^{\mathrm{tree}}_{j} + \hat{\mathcal{M}}^{\mathrm{loop}}_{j}
=
\mathcal{M}^{\mathrm{tree}}_{j} +
\mathcal{M}^{\mathrm{loop}}_{j} + \mathcal{M}^{\mathrm{CT}}_{j}.
\ee
Here,
$\mathcal{M}^{\mathrm{tree}}_{j}$ represents the tree-level (or LO) amplitude,
$\hat{\mathcal{M}}^{\mathrm{loop}}_{j}$ the renormalized one-loop amplitude,
$\mathcal{M}^{\mathrm{loop}}_{j}$ the non-renormalized one-loop amplitude and $\mathcal{M}^{\mathrm{CT}}_{j}$ the total counterterm.
The latter includes all the individual counterterms that end up contributing to the process $j$. Since different combinations of counterterms are taken as independent in the four $C_i$---in such a way that certain counterterms (like $\delta \alpha_0$) have different expressions in the different combinations---, $\mathcal{M}^{\mathrm{CT}}_{j}$ takes different values in the different combinations. In this way, $\hat{\mathcal{M}}_{j}$ depends on the combination $C_i$.

Once $\hat{\mathcal{M}}_{j}$ is calculated, one can determine the NLO decay width, $\Gamma^{\mathrm{NLO}}_{j}$. It is convenient to split the latter into three components:
\be
\Gamma^{\mathrm{NLO}}_{j}
=
\Gamma^{\mathrm{LO}}_{j}
+
\Gamma^{\mathrm{mix}}_{j}
+
\Gamma^{m_3}_{j}.
\label{eq:NLO-split}
\ee
On the right-hand side, 
the first term configures the pure LO contribution,
the second one represents the mixing between the tree-level and the renormalized one-loop contributions,
and the last one represents the contributions to the NLO decay width arising from the corrections to the mass of $h_3$.
The latter can be justified as follows.
In the calculation of the decay width of $j$, there will be occurences of the momentum squared of $h_3$. In principle, for an on-shell $h_3$, such quantity corresponds to the physical mass of $h_3$ squared, $m_{3\mathrm{P}}^2$. However, such identification would already contain corrections to a leading-order prediction, since $m_{3\mathrm{P}}^2$ includes NLO effects. Therefore, we calculate $\Gamma^{\mathrm{LO}}_{j}$ and $\Gamma^{\mathrm{mix}}_{j}$ assuming that the momentum squared of $h_3$ is $m_{3\textrm{R}}^2$, and we define $\Gamma^{m_3}_{j}$ as the NLO effects that show up when one calculates $\Gamma^{\mathrm{LO}}_{j}$ by associating the momentum squared of $h_3$ to $m_{3\mathrm{P}}^2$.%
\fn{
More precisely, $\Gamma^{m_3}_{j}$ is calculated by replacing all occurences of $m_{3\mathrm{R}}$ in $\Gamma^{\mathrm{LO}}_{j}$ by $m_{3\mathrm{R}}$ plus the NLO correction to this mass, and expanding in series to first order in the NLO correction. In this way, we avoid including several beyond-NLO effects that would show up through the cavalier replacement of $m_{3\mathrm{R}}$ by $m_{3\mathrm{P}}$. It should be clear, however, that the definition of the pole mass in eq. \ref{eq:m3P} already ignores beyond-NLO effects. It can be argued that such effects may also be ignored in the calculation of the decay width; this is the usual procedure in the Minimal Supersymmetric Standard Model (cf. e.g. \cite{
Heinemeyer:1996tg,
Brignole:1991pq,
Ellis:1991zd,
Brignole:1992zv,
Chankowski:1992es}).
Here, by calculating $\Gamma^{m_3}_{j}$ the way we do, not only do we obtain a more strict NLO calculation, but we also distinguish the NLO contributions arising from the corrections to the mass ($\Gamma^{m_3}_{j}$) from those arising from the corrections to the vertex ($\Gamma^{\mathrm{mix}}_{j}$).}
%

In what follows, we consider four different decays of $h_3$: 
to $ZZ$, $h_1 Z$, $h_2 Z$ and $h_2 h_1$.
For each of them, we write not only $\mathcal{M}^{\mathrm{tree}}_{j}$ and $\hat{\mathcal{M}}^{\mathrm{loop}}_{j}$ in terms of form factors, but also the three components of the right-hand side of eq. \ref{eq:NLO-split}.%
\fn{The form factors are identified by the letter \textit{f}, containing a subscript which starts with the particles in the final state, followed by a natural number. The attribution of natural numbers follows the conventions of  \FM \cite{Fontes:2019wqh,Fontes:2021iue}, which do not exploit momentum conservation.
Tree-level form factors are identified with the superscript `tree'. We omit the expressions corresponding to the form factors, which are in general extremely large.}

\subsubsection{$h_3 \to Z Z$}
\label{sec:h3ZZ-expressions}

For $h_3 \to Z Z$, we define the momenta and Lorentz indices such that $h_3(p_1) \to Z^{\nu}(q_1) Z^{\sigma}(q_2)$, so that:
\bs
\label{eq:ff-h3ZZ}
\bea
&&
\mathcal{M}_{h_3 \to ZZ}^{\mathrm{tree}}
= \varepsilon^*_{\nu}(q_1) \, \varepsilon^*_{\sigma}(q_2) \, \,  f_{\mathrm{Z}\mathrm{Z},3}^{\mathrm{tree}} \, g^{\nu\sigma}, \\[2mm]
&&
\hat{\mathcal{M}}_{h_3 \to ZZ}^{\mathrm{loop}}
= \varepsilon^*_{\nu}(q_1) \, \varepsilon^*_{\sigma}(q_2)
\bigg(
f_{\mathrm{Z}\mathrm{Z},3} \, g^{\nu\sigma} +
f_{\mathrm{Z}\mathrm{Z},6} \, p_1^{\nu} \, p_1^{\sigma} +
f_{\mathrm{Z}\mathrm{Z},9} \, p_1^{\nu} \, q_1^{\sigma} +
f_{\mathrm{Z}\mathrm{Z},24} \, p_1^{\sigma} \, q_2^{\nu} +
f_{\mathrm{Z}\mathrm{Z},27} \, q_1^{\sigma} \, q_2^{\nu} \nonumber\\[-3mm]
&&
\hs{25mm}
+ f_{\mathrm{Z}\mathrm{Z},33} \, p_1^{\omega} \, q_1^{\upsilon} \, \epsilon^{\nu\sigma\omega\upsilon} +
f_{\mathrm{Z}\mathrm{Z},15} \, p_1^{\sigma} \, q_1^{\nu} +
f_{\mathrm{Z}\mathrm{Z},18} \, q_1^{\nu} \, q_1^{\sigma} +
f_{\mathrm{Z}\mathrm{Z},21} \, q_1^{\nu} \, q_2^{\sigma}
\bigg).
\eea
\es
As a consequence,
\bs
\label{eq:h3-ZZ}
\begin{flalign}
\Gamma^{\mathrm{LO}}_{h_3 \to Z Z}
&=
\dfrac{\sqrt{m_{3\textrm{R}}^4 - 4 \, m_{3\textrm{R}}^2 \, m_{\mathrm{Z}}^2}}{128\, \pi \, m_{3\textrm{R}}^3 \, m_{\mathrm{Z}}^4} \left( m_{3\textrm{R}}^4 - 4 \, m_{3\textrm{R}}^2 \, m_{\mathrm{Z}}^2 + 12 \, m_{\mathrm{Z}}^4 \right) \left(f_3^{\mathrm{tree}}\right)^2, \\[3mm]
\Gamma^{\mathrm{mix}}_{h_3 \to Z Z}
&=
\dfrac{\sqrt{m_{3\textrm{R}}^4 - 4 \, m_{3\textrm{R}}^2 \, m_{\mathrm{Z}}^2}}{128\, \pi \, m_{3\textrm{R}}^3 \, m_{\mathrm{Z}}^4} f_3^{\mathrm{tree}}
\bigg\{ 2 \left( m_{3\textrm{R}}^4 - 4 \, m_{3\textrm{R}}^2 \, m_{\mathrm{Z}}^2 + 12 \, m_{\mathrm{Z}}^4 \right) \mathrm{Re} [f_3] \nonumber\\
& \hs{1mm}
+ m_{3\textrm{R}}^2 \,  \left( m_{3\textrm{R}}^4 - 6 \, m_{3\textrm{R}}^2 \, m_{\mathrm{Z}}^2 + 8 \, m_{\mathrm{Z}}^4 \right)  \, \operatorname{Re}[f_{\mathrm{Z}\mathrm{Z},6} + f_{\mathrm{Z}\mathrm{Z},9} + f_{\mathrm{Z}\mathrm{Z},24} + f_{\mathrm{Z}\mathrm{Z},27}] \bigg\}
, \\
\label{eq:Gm3-h3ZZ}
\Gamma^{m_3}_{h_3 \to Z Z}
&=
3 \, \, \Delta m_3 \, \, \Gamma^{\mathrm{LO}}_{h_3 \to Z Z} \, \,  
\dfrac{m_{3\mathrm{R}}^6 - 4 \, m_{3\mathrm{R}}^4 \, m_{\textrm{Z}}^2 - 4 \, m_{3\mathrm{R}}^2 \, m_{\textrm{Z}}^4 + 32 \, m_{\textrm{Z}}^6}{m_{3\mathrm{R}}^6 - 8 \, m_{3\mathrm{R}}^4 \, m_{\textrm{Z}}^2 + 28 \, m_{3\mathrm{R}}^2 \, m_{\textrm{Z}}^4 - 48 \, m_{\textrm{Z}}^6} .
\end{flalign}
\es

\subsubsection{$h_3 \to h_1 Z$}

Concerning the decay $h_3 \to h_1 Z$, we define the momenta and Lorentz indices such that $h_3(p_1) \to h_1(q_1) Z^{\sigma}(q_2)$, and we define form factors such that:
\be
\mathcal{M}_{h_3 \to h_1 Z}^{\mathrm{tree}}
=
\varepsilon^*_{\sigma}(q_2)
\Big(
f_{1\mathrm{Z},3}^{\mathrm{tree}} \, q_1^{\sigma} + f_{1\mathrm{Z},9}^{\mathrm{tree}} \, p_1^{\sigma}
\Big), \quad
\hat{\mathcal{M}}_{h_3 \to h_1 Z}^{\mathrm{loop}}
= \varepsilon^*_{\sigma}(q_2)
\Big(
f_{1\mathrm{Z},3} \, q_1^{\sigma} +
f_{1\mathrm{Z},6} \, q_2^{\sigma} +
f_{1\mathrm{Z},9} \, p_1^{\sigma}
\Big).
\ee
Then, we have:
\bs
\label{eq:h3-h1Z}
\begin{flalign}
\Gamma^{\mathrm{LO}}_{h_3 \to h_1 Z}
&=
\dfrac{
\left|f_{1\mathrm{Z},3}^{\mathrm{tree}} + f_{1\mathrm{Z},9}^{\mathrm{tree}}\right|^2
}{64 \, \pi \, m_{3\textrm{R}}^3 \, m_{\textrm{Z}}^2}
\Big(m_{3\textrm{R}}^4 + m_1^4 + m_{\textrm{Z}}^4 - 2 m_{3\textrm{R}}^2 m_1^2 - 2 m_{3\textrm{R}}^2 m_{\textrm{Z}}^2 - 2 m_1^2 m_{\textrm{Z}}^2 \Big)^{3/2}, \\[3mm]
\Gamma^{\mathrm{mix}}_{h_3 \to h_1 Z}
&=
\dfrac{
\mathrm{Re} \left[\left(f_{1\mathrm{Z},3}^{\mathrm{tree}} + f_{1\mathrm{Z},9}^{\mathrm{tree}}\right)
\left(f_{1\mathrm{Z},3}^* + f_{1\mathrm{Z},9}^*\right)\right]
}{32 \, \pi \, m_{3\textrm{R}}^3 \, m_{\textrm{Z}}^2}
\Big(m_{3\textrm{R}}^4 + m_1^4 + m_{\textrm{Z}}^4 - 2 m_{3\textrm{R}}^2 m_1^2 
\nonumber\\[-2mm]
& \hs{75mm}
- 2 m_{3\textrm{R}}^2 m_{\textrm{Z}}^2 - 2 m_1^2 m_{\textrm{Z}}^2 \Big)^{3/2}, \\
\label{eq:Gm3-h3h1Z}
\Gamma^{m_3}_{h_3 \to h_1 Z}
&=
-3 \, \, \Delta m_3 \, \, \Gamma^{\mathrm{LO}}_{h_3 \to h_1 Z} \, \, 
\dfrac{m_1^4 - m_{3\mathrm{R}}^4 - 2 \, m_1^2 \, m_{\textrm{Z}}^2 + m_{\textrm{Z}}^4 }{m_{3\textrm{R}}^4 + m_1^4 + m_{\textrm{Z}}^4 - 2 m_{3\textrm{R}}^2 m_1^2
- 2 m_{3\textrm{R}}^2 m_{\textrm{Z}}^2 - 2 m_1^2 m_{\textrm{Z}}^2}.
\end{flalign}
\es

\subsubsection{$h_3 \to h_2 Z$}

Similarly for $h_3 \to h_2 Z$, we define the momenta and Lorentz indices such that $h_3(p_1) \to h_2(q_1) Z^{\sigma}(q_2)$, and we define form factors such that:
\be
\mathcal{M}_{h_3 \to h_2 Z}^{\mathrm{tree}}
=
\varepsilon^*_{\sigma}(q_2)
\Big(
f_{2\mathrm{Z},3}^{\mathrm{tree}} \, q_1^{\sigma} + f_{2\mathrm{Z},9}^{\mathrm{tree}} \, p_1^{\sigma}
\Big), \quad
\hat{\mathcal{M}}_{h_3 \to h_2 Z}^{\mathrm{loop}}
= \varepsilon^*_{\sigma}(q_2)
\Big(
f_{2\mathrm{Z},3} \, q_1^{\sigma} +
f_{2\mathrm{Z},6} \, q_2^{\sigma} +
f_{2\mathrm{Z},9} \, p_1^{\sigma}
\Big).
\ee
Then,
\bs
\label{eq:h3-h2Z}
\begin{flalign}
\Gamma^{\mathrm{LO}}_{h_3 \to h_2 Z}
&=
\dfrac{
\left|f_{2\mathrm{Z},3}^{\mathrm{tree}} + f_{2\mathrm{Z},9}^{\mathrm{tree}}\right|^2
}{64 \, \pi \, m_{3\textrm{R}}^3 \, m_{\textrm{Z}}^2}
\Big(m_{3\textrm{R}}^4 + m_2^4 + m_{\textrm{Z}}^4 - 2 m_{3\textrm{R}}^2 m_2^2 - 2 m_{3\textrm{R}}^2 m_{\textrm{Z}}^2 - 2 m_2^2 m_{\textrm{Z}}^2 \Big)^{3/2}, \\[3mm]
\Gamma^{\mathrm{mix}}_{h_3 \to h_2 Z}
&=
\dfrac{\mathrm{Re} \left[
\left(f_{2\mathrm{Z},3}^{\mathrm{tree}} + f_{2\mathrm{Z},9}^{\mathrm{tree}}\right)
\left(f_{2\mathrm{Z},3}^* + f_{2\mathrm{Z},9}^*\right)\right]
}{32 \, \pi \, m_{3\textrm{R}}^3 \, m_{\textrm{Z}}^2}
\Big(m_{3\textrm{R}}^4 + m_2^4 + m_{\textrm{Z}}^4 - 2 m_{3\textrm{R}}^2 m_2^2
\nonumber\\[-2mm]
& \hs{75mm}
- 2 m_{3\textrm{R}}^2 m_{\textrm{Z}}^2 - 2 m_2^2 m_{\textrm{Z}}^2 \Big)^{3/2}, \\
\label{eq:Gm3-h3h2Z}
\Gamma^{m_3}_{h_3 \to h_2 Z}
&=
-3 \, \, \Delta m_3 \, \, \Gamma^{\mathrm{LO}}_{h_3 \to h_2 Z} \, \, 
\dfrac{m_2^4 - m_{3\mathrm{R}}^4 - 2 \, m_2^2 \, m_{\textrm{Z}}^2 + m_{\textrm{Z}}^4 }{m_{3\textrm{R}}^4 + m_2^4 + m_{\textrm{Z}}^4 - 2 m_{3\textrm{R}}^2 m_2^2 - 2 m_{3\textrm{R}}^2 m_{\textrm{Z}}^2 - 2 m_2^2 m_{\textrm{Z}}^2}.
\end{flalign}
\es

\subsubsection{$h_3 \to h_2 h_1$}
\label{sec:h3h2h1-expressions}

In this case, we simply have:
\be
\hat{\mathcal{M}}_{h_3 \to h_2 h_1}
=
\mathcal{M}_{h_3 \to h_2 h_1}^{\mathrm{tree}}
+ 
\hat{\mathcal{M}}_{h_3 \to h_2 h_1}^{\mathrm{loop}}
=
f_{21,1}^{\mathrm{tree}} + f_{21,1},
\ee
so that:
\bs
\label{eq:h3-h2h1}
\begin{flalign}
\Gamma^{\mathrm{LO}}_{h_3 \to h_2 h_1}
&=
\dfrac{\sqrt{m_{3\mathrm{R}}^4 + m_1^4 + m_2^4 - 2 m_{3\mathrm{R}}^2 m_1^2 - 2 m_{3\mathrm{R}}^2 m_2^2 - 2 m_1^2 m_2^2}}{16 \, \pi \, m_{3\textrm{R}}^3}
\left(f_{21,1}^{\mathrm{tree}}\right)^2, \\[3mm]
\Gamma^{\mathrm{mix}}_{h_3 \to h_2 h_1}
&=
\dfrac{\sqrt{m_{3\mathrm{R}}^4 + m_1^4 + m_2^4 - 2 m_{3\mathrm{R}}^2 m_1^2 - 2 m_{3\mathrm{R}}^2 m_2^2 - 2 m_1^2 m_2^2}}{8 \, \pi \, m_{3\textrm{R}}^3} \, \, f_{21,1}^{\mathrm{tree}} \, \, \textrm{Re} \left[f_{21,1}\right], \\[3mm]
\Gamma^{m_3}_{h_3 \to h_2 h_1}
&= - \Delta m_3 \, \, \Gamma^{\mathrm{LO}}_{h_3 \to h_2 h_1} \, \, \dfrac{3 \left( m_1^2 - m_2^2 \right)^2 - 4 \left( m_1^2 + m_2^2 \right)  \, m_{3\mathrm{R}}^2 + m_{3\mathrm{R}}^4}{m_{3\mathrm{R}}^4 + m_1^4 + m_2^4 - 2 m_{3\mathrm{R}}^2 m_1^2 - 2 m_{3\mathrm{R}}^2 m_2^2 - 2 m_1^2 m_2^2}.
\end{flalign}
\es

\section{Computational tools and simulation procedure}
\label{sec:procedure}

The Feynman rules, counterterms, one-loop amplitudes and decay widths were all calculated with \FM \cite{Fontes:2019wqh,Fontes:2021iue}, which resorts to \ts{FeynRules}~\cite{Christensen:2008py,Alloul:2013bka}, \ts{QGRAF}~\cite{Nogueira:1991ex} and \ts{FeynCalc}~\cite{Mertig:1990an,Shtabovenko:2016sxi,Shtabovenko:2020gxv}.
\FMS was also used to convert the results to \t{\ts{Fortran}}, where they were numerically evaluated using \ts{LoopTools}~\cite{Hahn:1998yk}.
For the scatter plots that follow, we generated points in the parameter space of the Type II C2HDM; we identified $h_1$ with the SM-like Higgs boson, implying $m_1 = 125 \, \, \textrm{GeV}$.
We restricted the parameter space by considering both theoretical and experimental constraints.
The former consist of
perturbative unitarity \cite{Akeroyd:2000wc,Ginzburg:2003fe},
boundeness from below \cite{Kanemura:1993hm},
the requirement of vacuum globality \cite{Ivanov:2015nea}
and the restrictions concerning the oblique parameters S, T, U \cite{Branco:2011iw}.
Regarding the experimental constraints, we demanded a $2 \sigma$ compatibility with the results coming from both $B \to X_s \gamma$ \cite{Deschamps:2009rh,Mahmoudi:2009zx,Hermann:2012fc,Misiak:2015xwa,Misiak:2017bgg} and $R_b$ \cite{Haber:1999zh,Deschamps:2009rh};
besides, signals for the SM-like Higgs boson were taken into account by requiring points to be compatible with the fits put forward in ref.
\cite{Aad:2019mbh}, whereas exclusion bounds coming from searches of extra Higgs bosons were included via \ts{HiggsBounds5} \cite{Bechtle:2020pkv}.
Our implementation of the calculation of the electric dipole moment (EDM) of the electron is based on refs. \cite{Abe:2013qla,Fontes:2015xva,Fontes:2017zfn}; 
we employed the (most severe) experimental current limit, $|d_e| < 1.1 \times 10^{-29} e \mbox{ cm}$ at $90\%$ confidence level, which was provided by the ACME collaboration \cite{Andreev:2018ayy}.
%
%
Finally, we choose as input parameters of the potential the set:
\be
\{p^V\}
=
\{\alpha_1, \, \alpha_2, \, \alpha_3, \, \beta, \, m_1, \, m_2, \, m_{\mathrm{H}^{+}}, \, \mathrm{Re} \, m_{12}^2
\},
\ee
where $m_{\mathrm{H}^{+}}$ is the mass of $H^+$,
%
and we vary them according to:
\begin{gather}
- \frac{\pi}{2} \le \alpha_{1,2,3} < \frac{\pi}{2},
\qquad
1 \le \tan \beta  \le 10,
\qquad
125 \mbox{ GeV } \leq m_2 < 800 \mbox{ GeV },
\no
580 \mbox{ GeV } \le m_{\mathrm{H}^\pm} <  800 \mbox{ GeV },
\qquad
0 \mbox{ GeV}^2 \le \mathrm{Re} \, m_{12}^2  < 500 000 \mbox{ GeV}^2.
\end{gather}
We generate and use three data sets, which optimize $\Gamma^{\mathrm{LO}}_{j}$ for $j = h_3 \to ZZ, h_3 \to h_1Z$ and $h_3 \to h_2Z$, respectively.%
\fn{Concerning $h_3 \to h_2 h_1$, see discussion in section \ref{sec:h3h2h1}.}
This separation is relevant, since different regions of parameter space lead to 
non-negligible values of $\Gamma^{\mathrm{LO}}_{j}$ according to the decay $j$, as shall be discussed in the following section.

\section{Results and discussion}
\label{sec:results}

\subsection{The mass of $h_3$}
\label{sec:results-mass}

\begin{figure}[h!]
\centering
\subfloat{\includegraphics[width=0.49\linewidth]{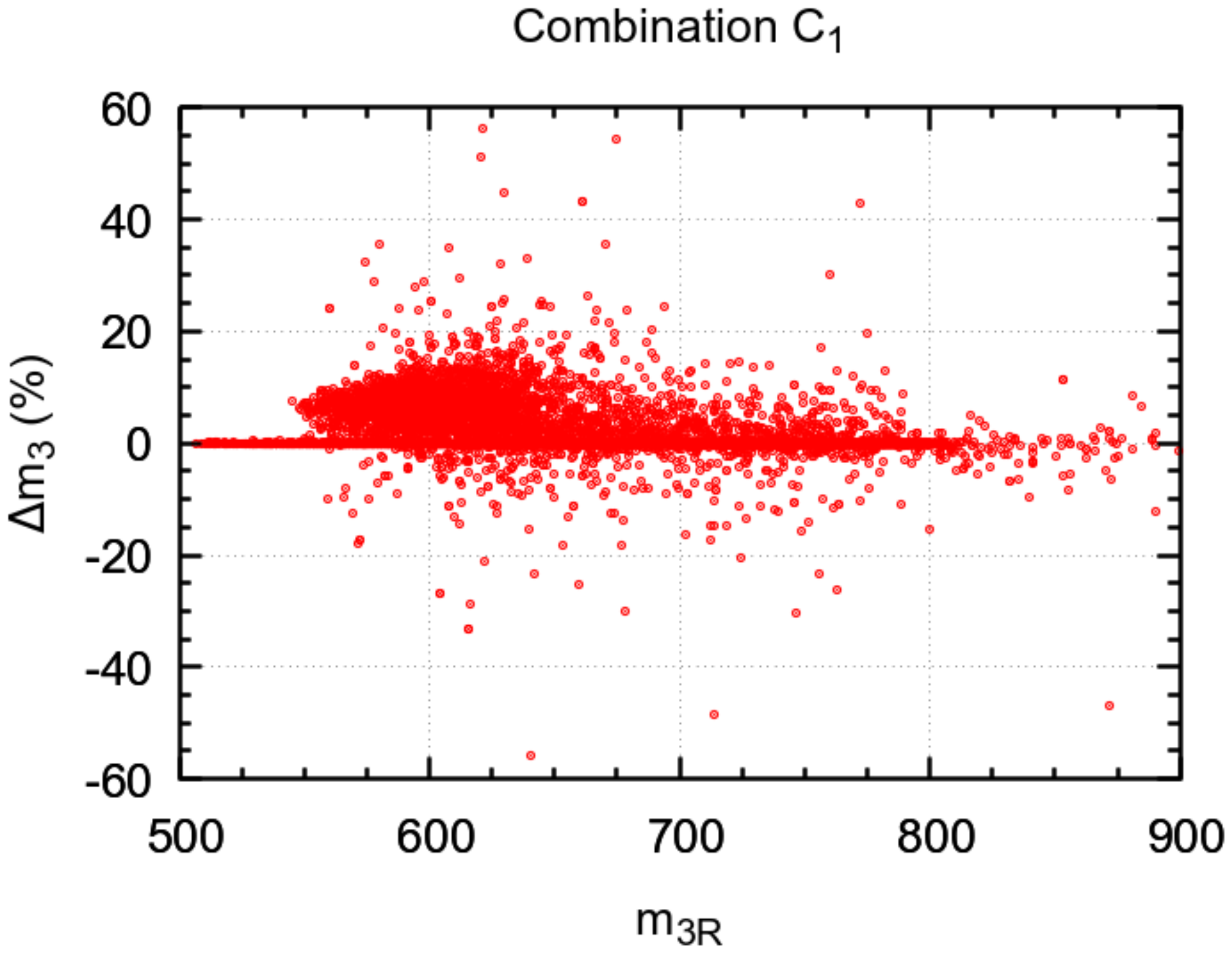}}
\ \
\subfloat{\includegraphics[width=0.49\linewidth]{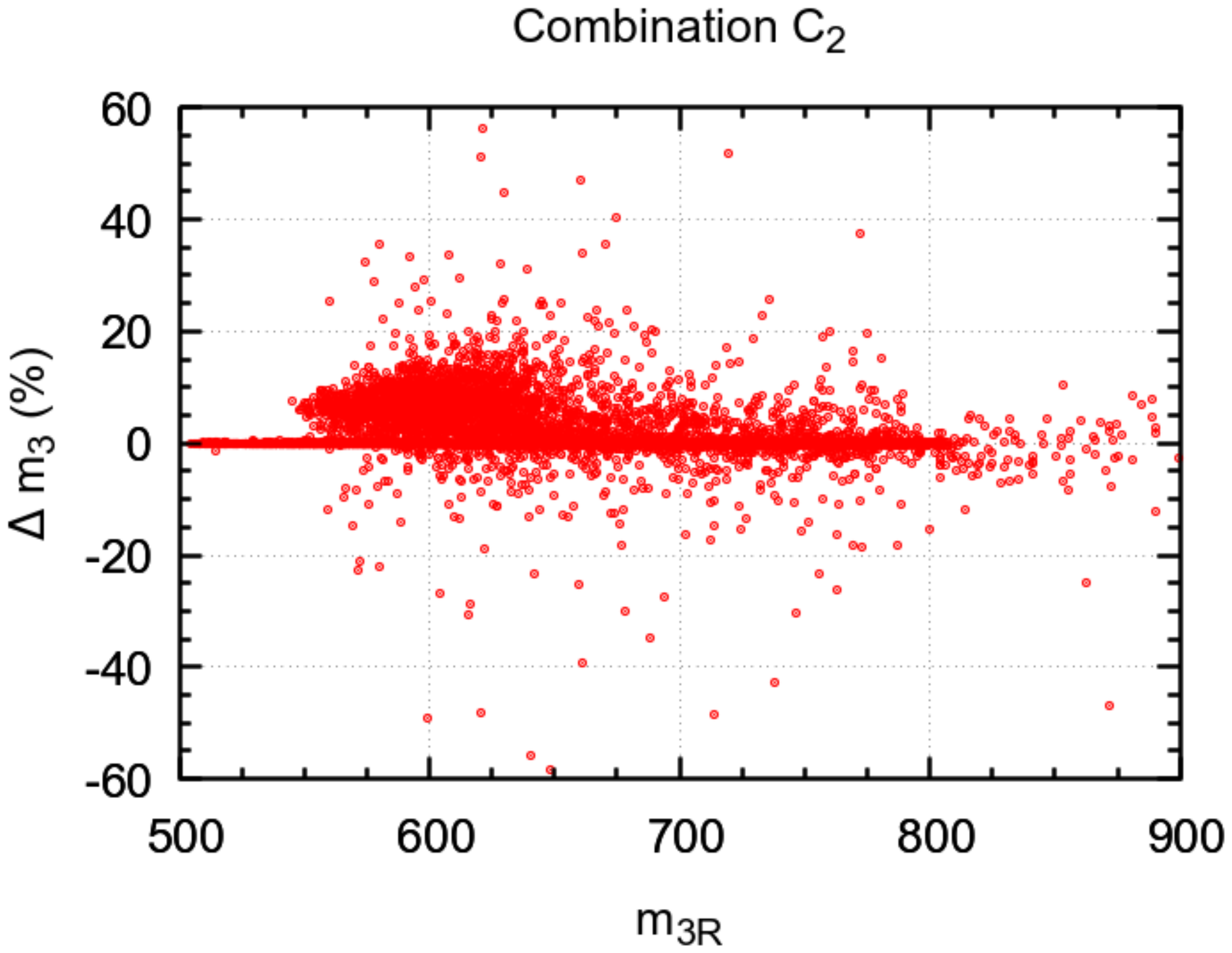}}\\
\subfloat{\includegraphics[width=0.49\linewidth]{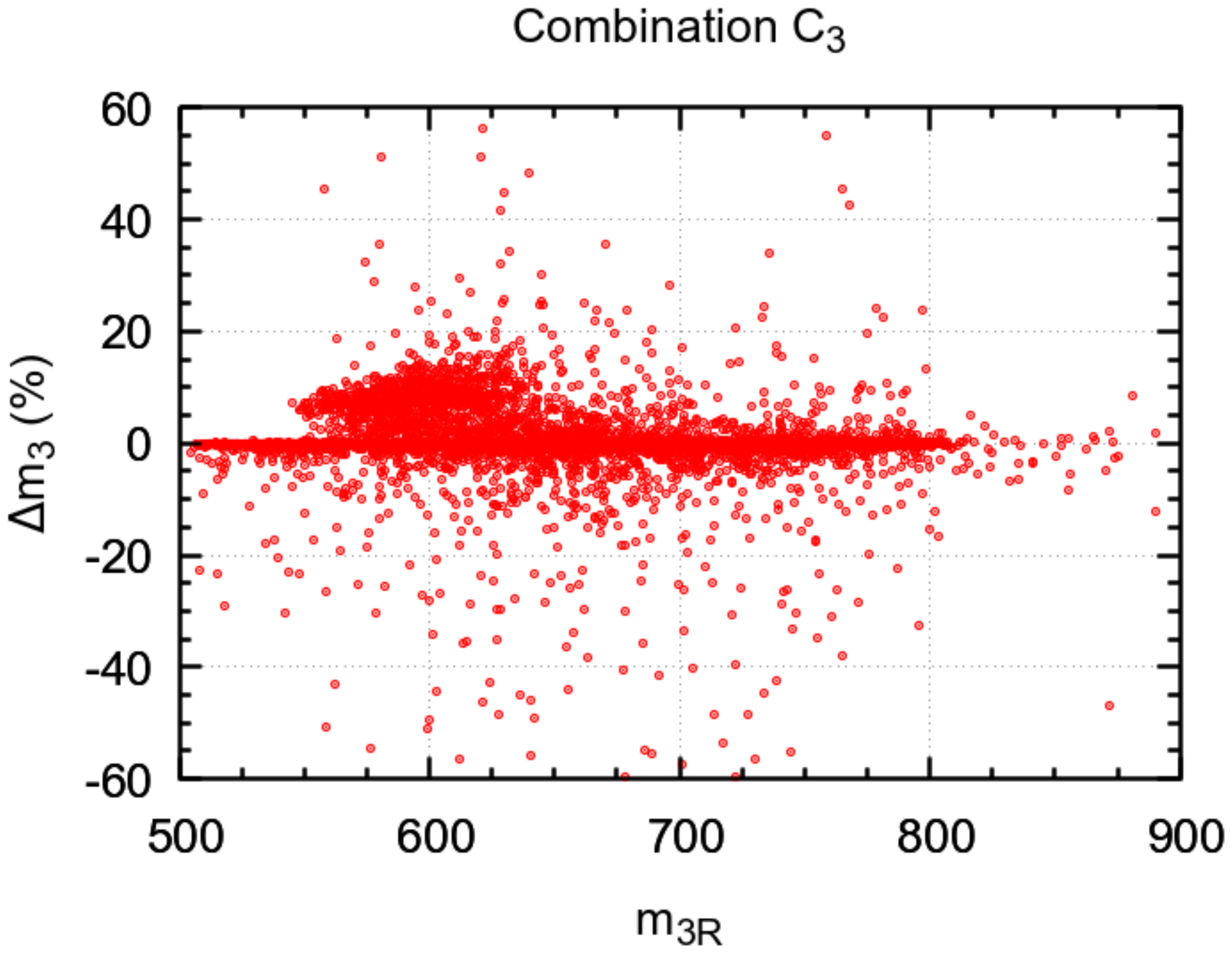}}
\ \
\subfloat{\includegraphics[width=0.49\linewidth]{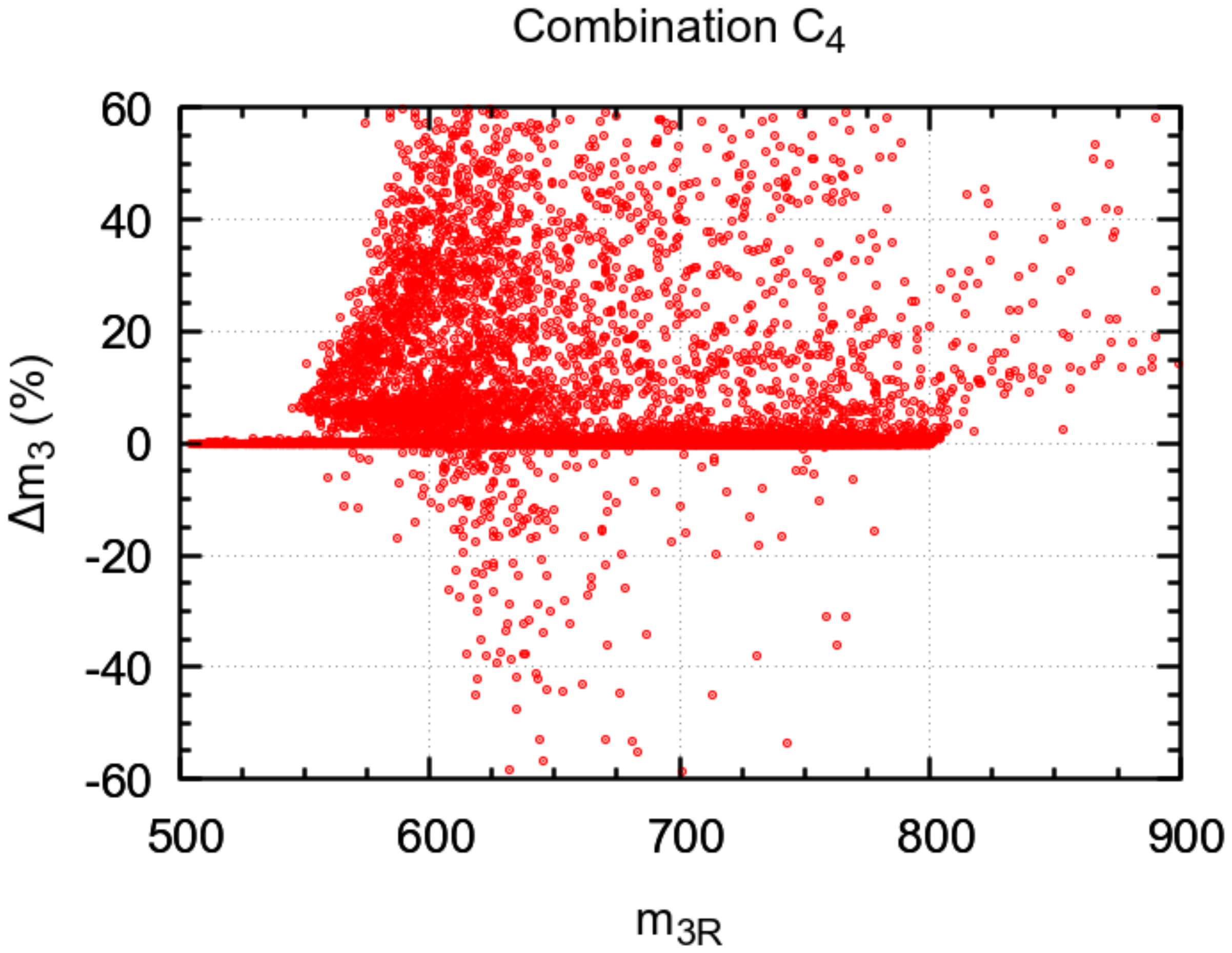}}
\caption{The quantity $\Delta m_3$ defined in eq. \ref{eq:Deltam3} for the different combinations: $C_1$ (top left), $C_2$ (top right), $C_3$ (bottom left) and $C_4$ with $\mu_{\mathrm{R}}=$ 350 GeV (bottom right). Only the range $-60\% < \Delta m_3 < 60\%$ is shown.}
\label{fig:Delta-m3-original}
\end{figure}

We start by investigating $\Delta m_3$ (defined in eq. \ref{eq:Deltam3}) in each of the four combinations $C_i$.%
\fn{As mentioned above, the combination $C_4$ has an explicit dependence on the renormalization scale, $\mu_{\mathrm{R}}$. In ref. \cite{Fontes:2021iue}, we studied three different scales for different processes calculated in $C_4$, and concluded that $\mu_{\mathrm{R}} = 350$ GeV led to the most well-behaved results. In what follows, we also use this scale.}
The results are shown in figure \ref{fig:Delta-m3-original}, as a function of the renormalized mass.
All the points coming from the three data sets mentioned in the last section are superimposed in the figure.
It is clear that the four combinations contain points which describe very large corrections. The reason can be traced back to $\delta m_3^2$, which contributes to $m_{3\mathrm{P}}$ (and thus to $\Delta m_3$) according to eq. \ref{eq:m3P}. As it turns out, $\delta m_3^2$ takes very large values in all combinations for very specific regions of the parameter space. To understand this aspect, notice that eq. \ref{eq:m3_derived} can be rewritten as:
\be
m_{3\mathrm{R}}^2 = \frac{m_1^2 \, \epsilon_1 + m_2^2 \, \epsilon_2}{(-\epsilon_3)},
\label{eq:m32_var}
\ee
where we define
\be
\epsilon_k \equiv R_{k3} \left( R_{k2} \tan \beta - R_{k1} \right),
\ee
where $k =  \{1,2,3\}$ and where the matrix $R$ is defined in eq. \ref{eq:matrixR}.
We are interested in the case where $\epsilon_3$ takes very small values, approaching zero. Since the expression for $\epsilon_3$ is:
\be
\epsilon_3 = 
c_2 \, c_3 \,
\bigg\{
c_1 \Big[s_2 c_3 - s_3 \tan (\beta) \Big]
- s_1 \Big[s_2 c_3 \tan (\beta) + s_3 \Big]
\bigg\},
\ee
we can distinguish two different scenarios---identified as $S_1$ and $S_2$ in the following---that lead to a vanishing $\epsilon_3$:%
\fn{In principle, $\alpha_2 = \pm \dfrac{\pi}{2}$ could also lead to a vanishing $\epsilon_3$. However, the theoretical constraints of the theory (in particular, those related to the electron EDM) force $\alpha_2$ to take very small values.}
\be
S_1: \alpha_3 = \pm \dfrac{\pi}{2},
\hspace{20mm}
S_2: \alpha_3 = \arctan \Big[\cot(\alpha_1 + \beta) \sin(\alpha_2)\Big].
\label{eq:scenarios}
\ee
We thus see that, for $\alpha_3$ close to either one or the other limit, $m_{3\mathrm{R}}^2$ will tend to assume very large values, due to the smallness of the denominator $\epsilon_3$. There is, however, an exception: even if $\alpha_3$ is close either to $S_1$ or $S_2$, large values of $m_{3\mathrm{R}}^2$ can be avoided if $\epsilon_1$ and $\epsilon_2$ are simultaneously small. In fact, if we happen to choose points in a very particular region of the parameter space where not only $\epsilon_3$, but also $\epsilon_1$ and $\epsilon_2$ approach zero, then the smallness of the latter ($\epsilon_1$ and $\epsilon_2$) compensates for the smallness of the former ($\epsilon_3$), so that $m_{3\mathrm{R}}^2$ ends up taking moderate values.
Notice that \textit{im}moderate values $m_{3\mathrm{R}}^2$ would violate the theoretical constraints of the theory.%
\fn{More specifically, they would violate the oblique parameter T.
Note that the constraints are applied to the leading-order quantities; hence, the quantity of relevance here is $m_{3\mathrm{R}}^2$, not $m_{3\mathrm{P}}^2$.}
As a consequence, if $\epsilon_3$ is very small (through either $S_1$ or $S_2$), valid points can only be obtained if they are in the very fine-tuned region of parameter space which we alluded to (where both $\epsilon_1$ and $\epsilon_2$ approach zero) and which ensures that $m_{3\mathrm{R}}^2$ is not very large.

It turns out that, in such particular region, there is a problem with the counterterm $\delta m_3^2$, as we now explain. At tree-level, and like we saw at the end of section \ref{sec:model}, the mass of $h_3$ is a derived parameter, which takes different (complicated) expressions in the different combinations. When we go up to one-loop level, each of those expressions (meanwhile identified as bare) generates a renormalized term and a counterterm, according to eq. \ref{eq:mass-scalar-expa-2}. This happens in such a way that, as described in section \ref{sec:the-mass-ini}, the renormalized term is given by eq. \ref{eq:m3_derived} in all the four combinations, whereas the counterterm $\delta m_3^2$ depends on the combination $C_i$. Now, for each combination, the expression for $\delta m_3^2$ is closely related to eq. \ref{eq:m3_derived}, in the sense that both derive from the same original, bare expression. This suggests that, whenever eq. \ref{eq:m3_derived} tends to be very large, the values for the counterterm $\delta m_3^2$ in the different combinations will follow that trend. In particular, both scenarios described in eq. \ref{eq:scenarios} will naturally lead to very large values of $\delta m_3^2$. However, because the expressions for $\delta m_3^2$ are different from that of $m_{3\mathrm{R}}^2$ (although related to it, as we saw), the very precise fine tuning that leads the latter to adquire a moderate value turns out not to be verified in the former. We thus end up with a peculiar situation: the points (not excluded by theoretical constraints) obeying the scenarios \ref{eq:scenarios} require a very precise fine-tuning; this fine-tuning ensures that eq. \ref{eq:m3_derived} has moderate values (otherwise, the points would be exluded),
but does \textit{not} ensure that the expressions for $\delta m_3^2$ in the different combinations are also moderate. Hence, those points are valid points---since they pass all the constraints described in section \ref{sec:procedure}---, but lead to very large values for the counterterm $\delta m_3^2$ in the different combinations. 

In sum, there are points in very precise regions of the parameter space of the model that, although leading to acceptable values of $m_{3\mathrm{R}}^2$ (due to a razor-sharp fine-tuning), generate very large values of the counterterm $\delta m_3^2$; the regions of parameter space at stake are those around the scenarios $S_1$ and $S_2$ described in eq. \ref{eq:scenarios}. Three notes are in order here.

First, as we are about to see in detail, the precise regions at stake are indeed very precise (i.e. very fine-tuned), so that they are prone to be avoided by a random generation of points.
Second, the immoderate values of $\delta m_3^2$ do \textit{not} stem from an inappropriate choice of subtraction schemes. It does not happen, indeed, that the finite parts of the independent counterterms that $\delta m_3^2$ depends on become unusually large for the points at stake. Rather, the problem lies in the fact that the counterterms $\delta m_3^2$ in the different combinations correspond to complicated expressions,
in such manner that those expressions diverge in certain particular limits of the renormalized parameters. The complicated character of the counterterms $\delta m_3^2$, in turn, results from the circumstance that the bare parameter $m_{3(0)}^2$ depends in a complicated way on the independent parameters.%
\fn{One can argue that it is not really necessary to take the mass of $h_3$ as a dependent parameter of the C2HDM; indeed, although taking it as a dependent parameter is the most common option (see e.g. refs. \cite{ElKaffas:2006gdt,
Osland:2008aw,Arhrib:2010ju,Barroso:2012wz,Fontes:2017zfn}), one could also select one of the mixing angles instead \cite{Inoue:2014nva,Chen:2015gaa}.
We follow here ref. \cite{Fontes:2021iue}, which proposed the renormalization of the model assuming the mass of $h_3$ as a derived parameter; the exploration of an alternative renormalization---where a mixing angle is taken as dependent instead of the mass of $h_3$---is beyond the scope of this paper.}
Finally, a situation like the one described here---where, for certain regions of the parameter space, a dependent counterterm (and particularly a mass counterterm) takes large values due to the expression through which it depends on the independent counterterms---is common in models such as the Minimal Symmetric Standard Model \cite{Baro:2009gn,Heinemeyer:2010mm}.

\begin{figure}[h!]
\centering
\subfloat{\includegraphics[width=0.49\linewidth]{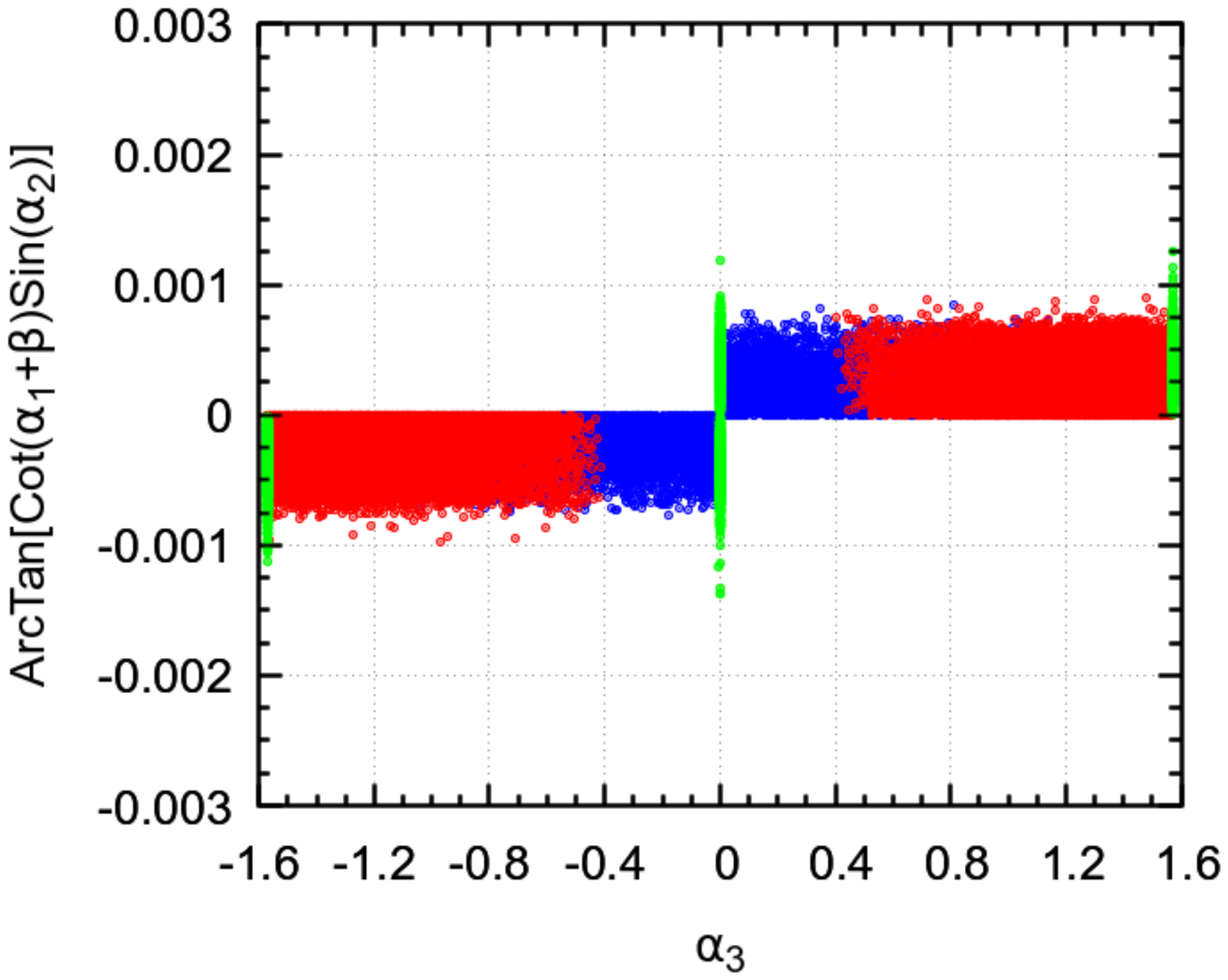}}
\ \
\subfloat{\includegraphics[width=0.49\linewidth]{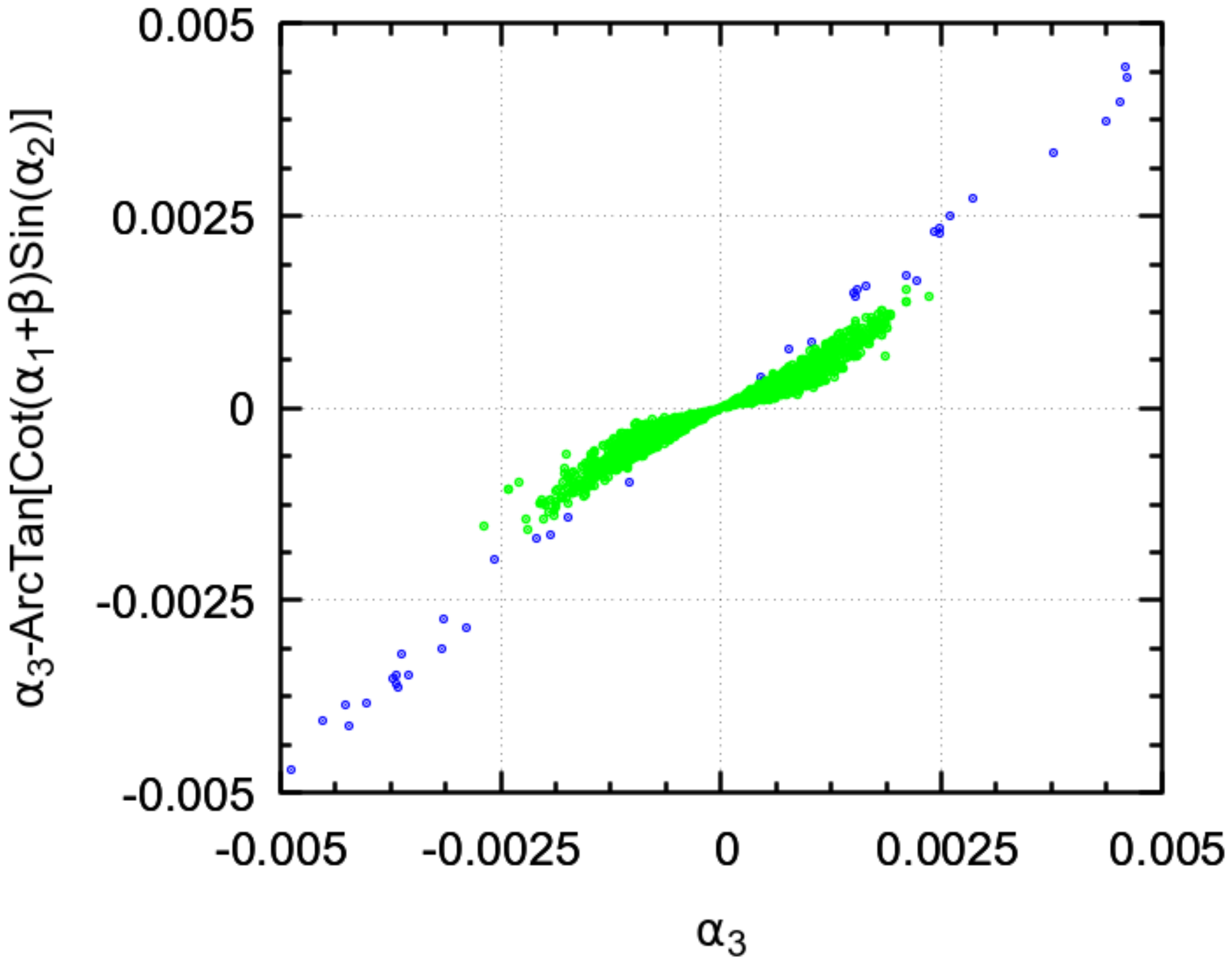}}\\
\caption{
Left panel: $\arctan \big[\cot(\alpha_1 + \beta) \sin(\alpha_2)\big]$ against $\alpha_3$, for the whole range of $\alpha_3$.
Right panel: the difference between $\alpha_3$ and $\arctan \big[\cot(\alpha_1 + \beta) \sin(\alpha_2)\big]$ against $\alpha_3$, for a very restricted range of $\alpha_3$.
In red, points leading to a non-negligible decay width for $h_3 \to ZZ$;
in blue, points leading to a non-negligible decay width for $h_3 \to h_1Z$;
in green, points allowing the decay $h_3 \to h_2 Z$.
}
\label{fig:S2}
\end{figure}

In the following, we want to avoid points in the aforementioned very fine-tuned regions---i.e. points very near to either $S_1$ or $S_2$. To better grasp the scenario $S_2$, we ascertain in figure \ref{fig:S2} the region of parameter space that it covers.
In both panels, we represent
in red a set of points optimized for the decay $h_3 \to ZZ$,
in blue a set of points optimized for the decay $h_3 \to h_1Z$
and in green a set of points which allow the process $h_3 \to h_2 Z$ at tree-level (by complying to $m_{3\mathrm{R}} > m_2 + m_{\mathrm{Z}}$).
From the red and the blue points on the left plot, it is clear that $\arctan \big[\cot(\alpha_1 + \beta) \sin(\alpha_2)\big]$ always takes very small values for the whole set of points; therefore, the scenario $S_2$ will be verified if and only if  $\alpha_3$ is very close to zero. This conclusion is also derived from the blue points on the right plot, where we show the difference between $\alpha_3$ and $\arctan \big[\cot(\alpha_1 + \beta) \sin(\alpha_2)\big]$ against $\alpha_3$, for a very restricted range of the latter around zero. In sum: in order to avoid $S_2$, points with $\alpha_3 \simeq 0$ must be avoided.

It is only after reaching such conclusion that we can properly appreciate the points represented in green. From the left plot, we see that those points (which, recall, are the ones enabling the decay $h_3 \to h_2 Z$) are restricted to very fine-tuned regions. More than just that: those very fine-tuned regions are \textit{precisely} those around the troublesome scenarios $S_1$ and $S_2$; indeed, they are concentrated either around $\alpha_3 = \pm \pi/2$ (scenario $S_1$) or around $\alpha_3 = 0$ (scenario $S_2$). 
We can thus antecipate a tension which we will come back to in section \ref{section:h3h2Z} below: in the decay $h_3 \to h_2 Z$, since all the points allowing it are very close to either $S_1$ or $S_2$, the values of $\delta m_3^2$ are generally very large, which in turn implies very large values of $\Gamma^{\mathrm{NLO}}_{h_3 \to h_2 Z}$ (recall eqs. \ref{eq:m3P}, \ref{eq:Deltam3}, \ref{eq:NLO-split} and \ref{eq:Gm3-h3h2Z}). So, if one tries to avoid the regions $S_1$ and $S_2$---in an attempt to obtain more stable results---, one is left with no points. In that decay, therefore, a compromise must be found: on the hand, one may wish to avoid the troublesome regions where $\delta m_3^2$ has a bad behaviour; on the other hand, one cannot significantly depart from those regions, for otherwise one ends up having no points. Notice that such compromise does not exist in the decays $h_3 \to ZZ$ and $h_3 \to h_2 Z$, since one can find points which allow these processes and which are far away from both $S_1$ and $S_2$.%
\fn{The decay $h_3 \to h_2 h_1$ suffers from the same problem as $h_3 \to h_2 Z$, and will be considered in detail in section \ref{sec:h3h2h1}.}

\begin{figure}[h!]
\subfloat{\includegraphics[width=0.49\linewidth]{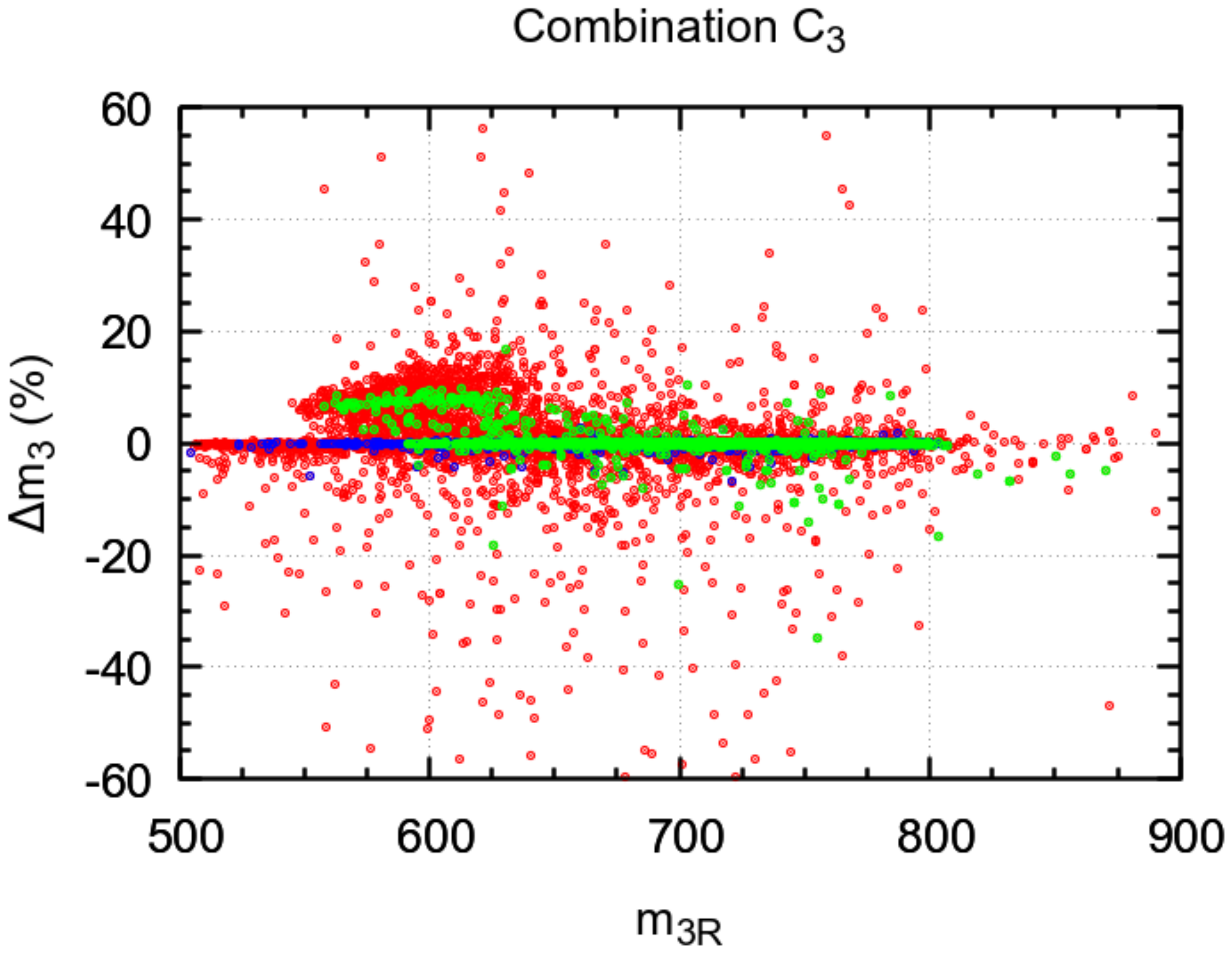}}
\ \
\subfloat{\includegraphics[width=0.49\linewidth]{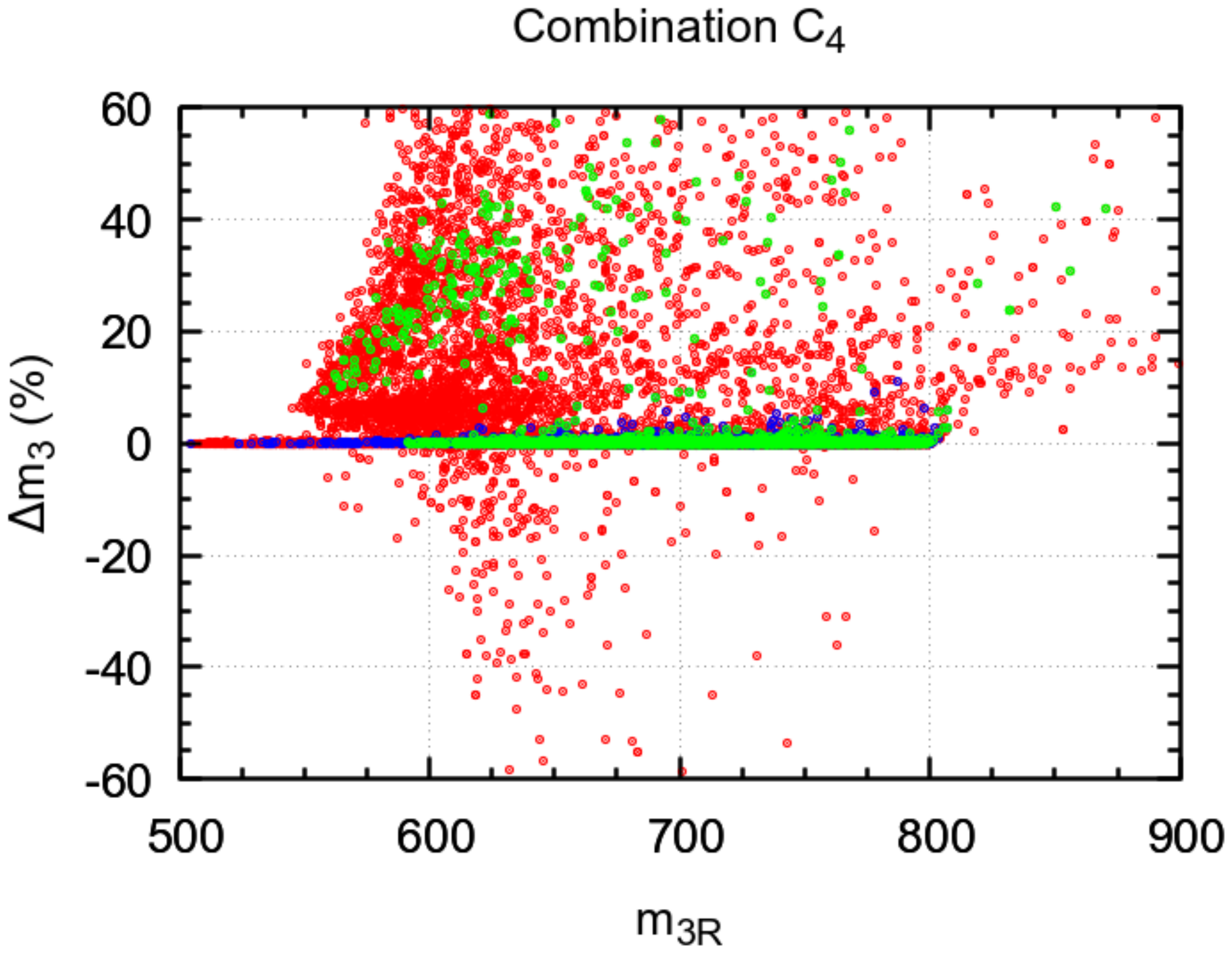}}
\caption{The same as in figure \ref{fig:Delta-m3-original}, but with extra cuts:
in $C_1$,
blue points have $\pi/2 -|\alpha_3| > 0.0005$ and $|\alpha_3|> 0.0005$, whereas
green points have $\pi/2 -|\alpha_3| > 0.0007$ and $|\alpha_3|> 0.00085$;
in $C_2$,
blue points have $\pi/2 -|\alpha_3| > 0.0005$ and $|\alpha_3|> 0.0005$, whereas
green points have $\pi/2 -|\alpha_3| > 0.00075$, $|\alpha_3|> 0.001075$ and $|\alpha_2|> 0.001$;
in $C_3$,
blue points have $\pi/2 -|\alpha_3| > 0.005$, $|\alpha_3|> 0.05$ and $|\alpha_2|> 0.0025$, whereas
green points have $\pi/2 -|\alpha_3| > 0.0004$ , $|\alpha_3|> 0.005$ and  $|\alpha_2|> 0.005$;
$C_4$ is as $C_3$.}
\label{fig:Delta-m3-corrected}
\end{figure}

This difference between the processes---as well as the differences between the combina- tions---leads us to consider different types of cuts, in our attempt to avoid the troublesome regions $S_1$ and $S_2$. This is illustrated in figure \ref{fig:Delta-m3-corrected},
\begin{figure*}[h!]
\centering
\subfloat{\includegraphics[width=0.49\linewidth]{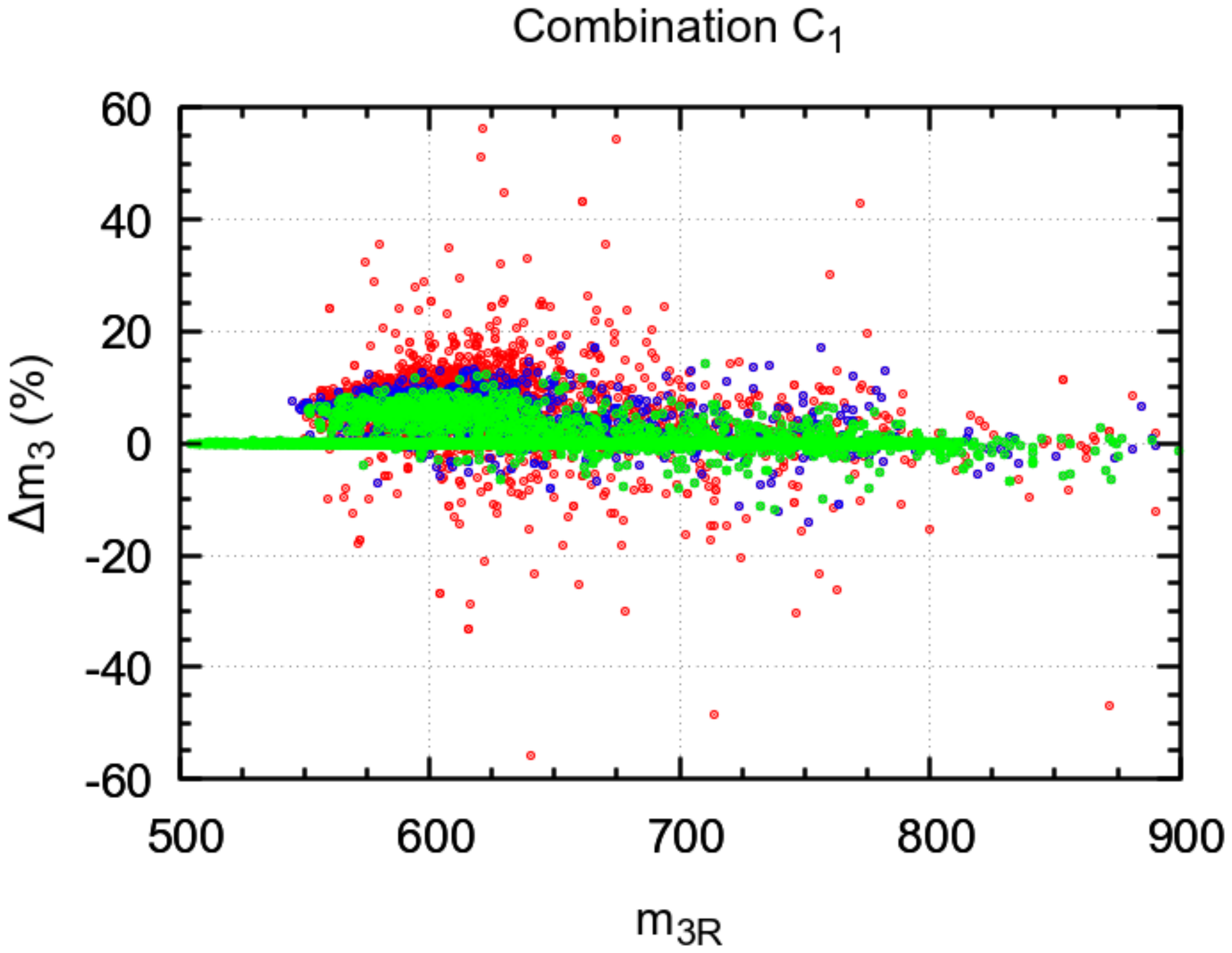}}
\ \
\subfloat{\includegraphics[width=0.49\linewidth]{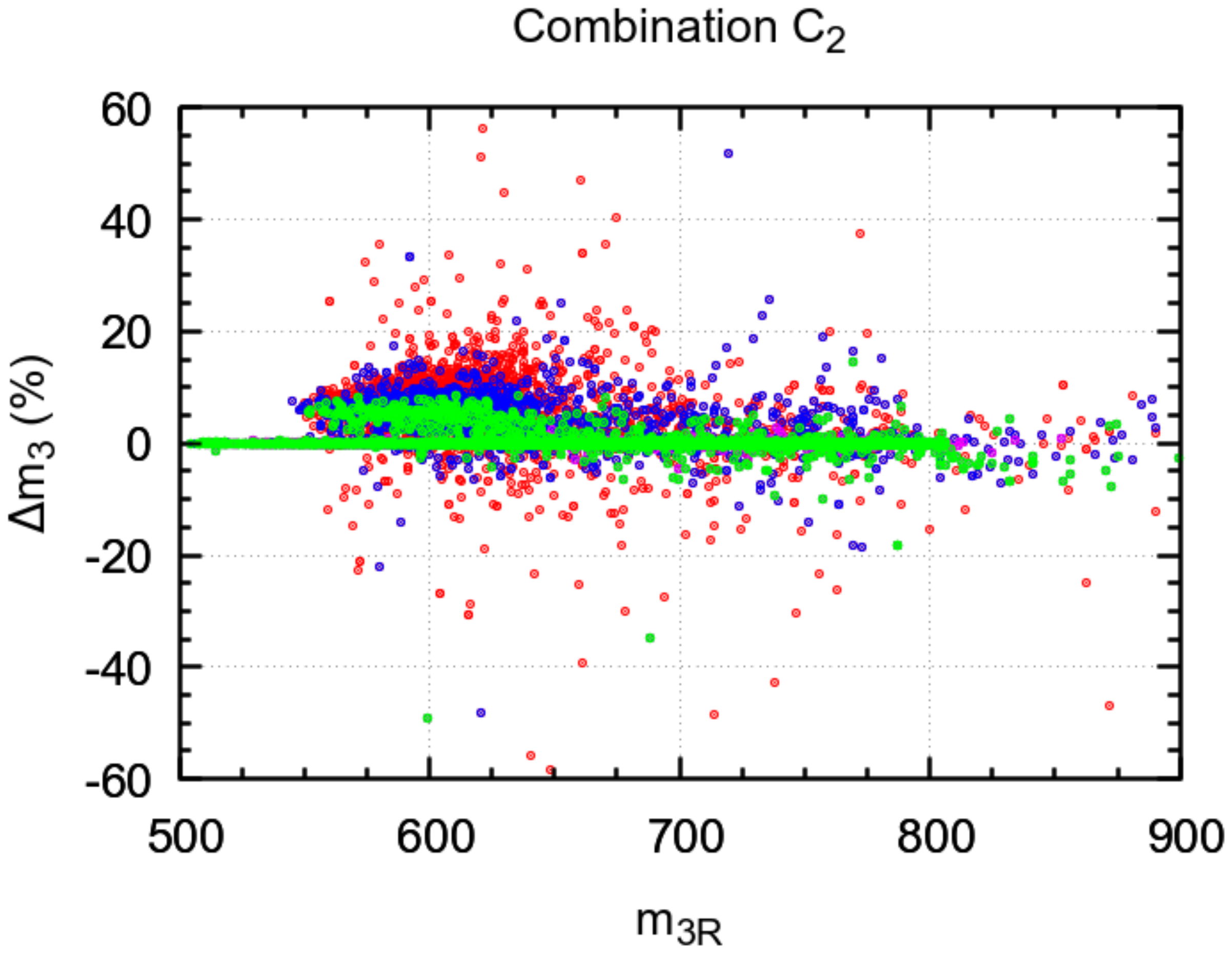}}
\end{figure*}
where we show the same as in figure \ref{fig:Delta-m3-original}, but now including different cuts on the mixing angles.
In all combinations, the points in blue correspond to cuts that shall be sufficient to generate adequate results for the decay $h_3 \to h_1 Z$, whereas the points in green correspond to cuts more adapted to $h_3 \to h_2 Z$.%
\fn{As for the remaning processes discussed in this paper: $h_3 \to Z Z$ will require cuts less stringent than the ones represented in blue in figure \ref{fig:Delta-m3-corrected}, while $h_3 \to h_2 h_1$ will not have a sufficiently relevant branching ratio to justify a detailed analysis (for details, see section \ref{sec:h3h2h1} below).}
Some aspects should be highlighted here.

First, it is clear from all the plots that the more stringent the cuts, the smaller the range of $\Delta m_3$. Or, which is the same, the more the scenarios $S_1$ and $S_2$ are avoided, the better the numerical stability of the corrections to the mass of $h_3$. This proves that the numerical instabilities observed in the red points have to do with a proximity of the points to either $S_1$ or $S_2$, so that the instabilities disappear when those scenarios are avoided.
Second, even after the cuts, we can still find points leading to large instabilities. This is especially true in the case of the blue points in the combination $C_2$, where values as large as $\Delta m_3 \simeq 50\%$ can still be found. However, this is only because the cuts at stake are not sufficiently stringent; for example, in combination $C_2$, the green points---which involve more severe cuts---lead to more stable results than the blue points. 
%
Finally, the combination $C_4$---which explicitly depends on the renormalization scale $\mu_{\mathrm{R}}$---is shown for $\mu_{\mathrm{R}}=$ 350 GeV. Yet, we found that the results strongly depend on the value of $\mu_{\mathrm{R}}$. For that reason, it will not be considered in what follows.

\subsection{NLO decay widths of $h_3$}

We now turn to the numerical results for the NLO decay widths. We want to investigate the importance of the NLO contributions. We thus define the relative correction to the decay width of $j$ as:
\be
\Delta \Gamma_j  \equiv  \dfrac{\Gamma^{\mathrm{NLO}}_j - \Gamma^{\mathrm{LO}}_j}{\Gamma^{\mathrm{LO}}_j}
=
\dfrac{\Gamma^{\mathrm{mix}}_{j} + \Gamma^{m_3}_{j}}{\Gamma^{\mathrm{LO}}_j},
\label{eq:Gamma-def}
\ee
where we used eq. \ref{eq:NLO-split}.
Notice that both $\Gamma^{\mathrm{mix}}_{j}$ and $\Gamma^{m_3}_{j}$ depend on the combination $C_i$;
the contribution coming from the former is expected to be \textit{grosso modo} similar to that found in the decays of $h_2$ studied in ref. \cite{Fontes:2021iue}.%
\fn{In fact, for a certain pair of final particles, the independent parameter counterterms are the same in the decays of $h_2$ and $h_3$, and the field counterterms (which are different in the decays of $h_2$ and $h_3$) are all fixed through the same subtraction scheme; cf. ref. \cite{Fontes:2021iue} for details.}
Here, we are particularly interested in the contribution coming from $\Gamma^{m_3}_{j}$ (which was absent in decays of $h_2$, since $m_2$ is an independent parameter).
From the expressions presented in section \ref{sec:h3ZZ-expressions} to \ref{sec:h3h2h1-expressions}, we see that, for every process $j$, a non-null difference between the pole mass and the renormalized mass of $h_3$ (i.e. $\Delta m_3 \neq 0$) leads to a non-null $\Gamma^{m_3}_{j}$.%
\fn{This happens despite the fact that there are no corrections coming from wave-function renormalization factors (also known as LSZ factors). Indeed, the circumstance that the field counterterms in eq. \ref{eq:neutral-field-CTs} were all fixed in the OSS scheme means that the wave-function renormalization factors become trivial \cite{Fontes:2021kue}. Nonetheless, because the mass of $h_3$ is a derived parameter, there are in general contributions to the NLO decay width arising from their loop corrections, as explained in section \ref{section:decays-expressions}.}
Actually, $\Gamma^{m_3}_{j}$ depends linearly on $\Delta m_3$; hence, the larger is $\Delta m_3$, the larger is $\Gamma^{m_3}_{j}$, and thus the larger is $\Delta \Gamma_j$. If face of this, we we are now able to better grasp the consequences of the plots in figure \ref{fig:Delta-m3-corrected}; in fact, we realize that a large value of $\Delta m_3$ will generally imply large corrections for two types of observables: not only that corresponding to the physical mass of $h_3$, but will to those corresponding to the decay widths of $h_3$ decays.

We now present results for the four decays described in section \ref{section:decays-expressions}, following the same order.

\subsubsection{$h_3 \to Z Z$}

\begin{figure}[h!]
\centering
\subfloat{\includegraphics[width=0.55\linewidth]{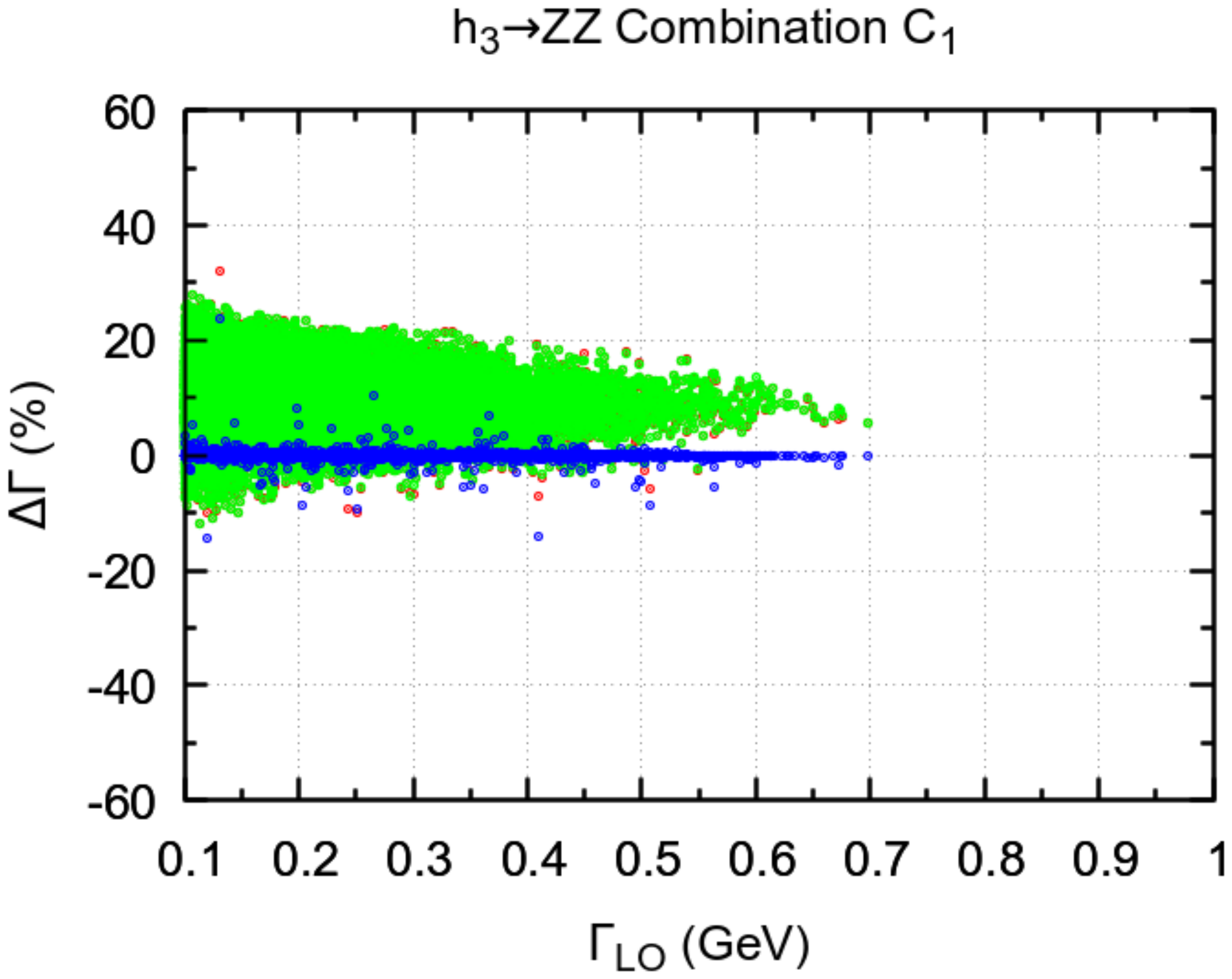}}\\
\subfloat{\includegraphics[width=0.49\linewidth]{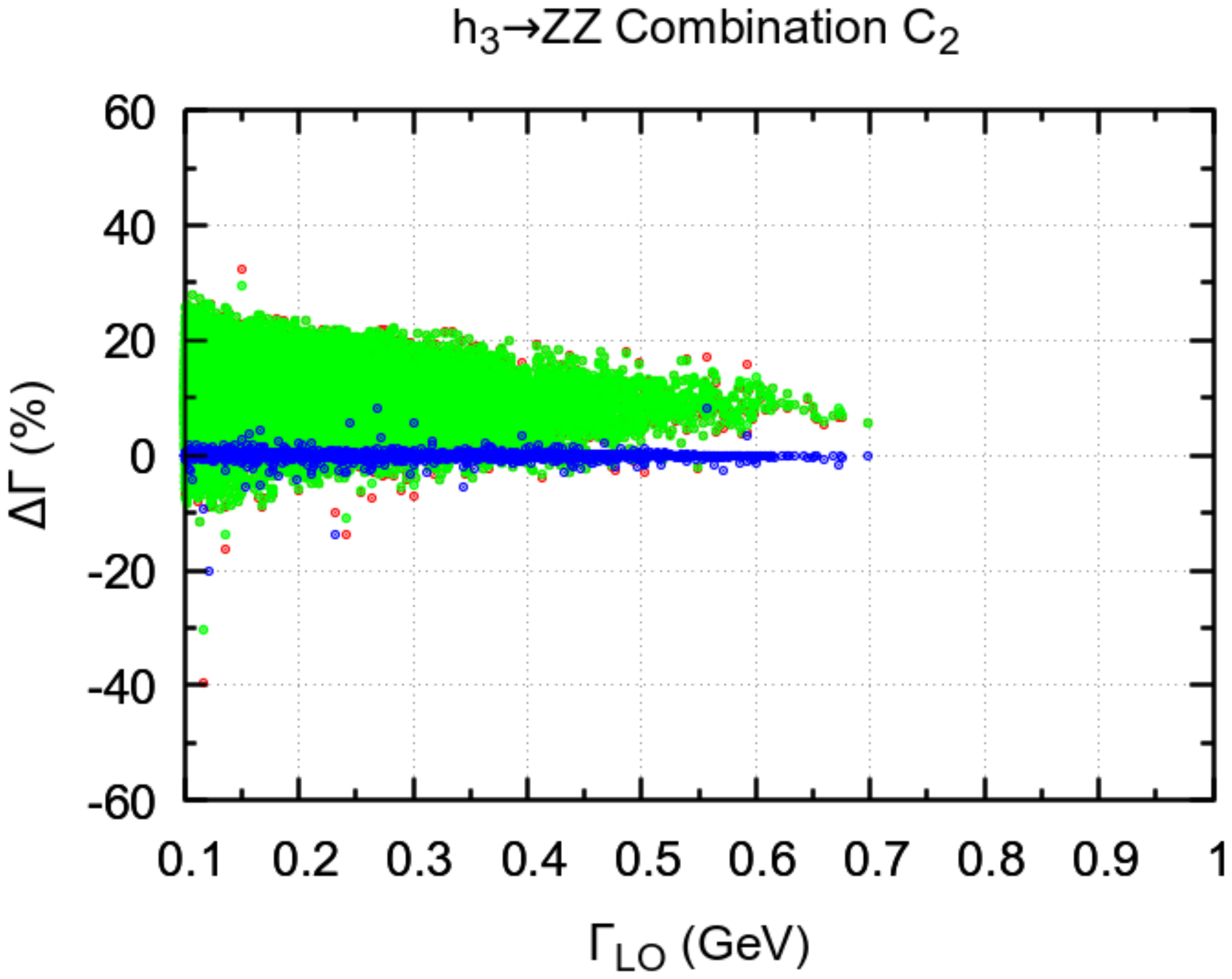}}\ \
\subfloat{\includegraphics[width=0.49\linewidth]{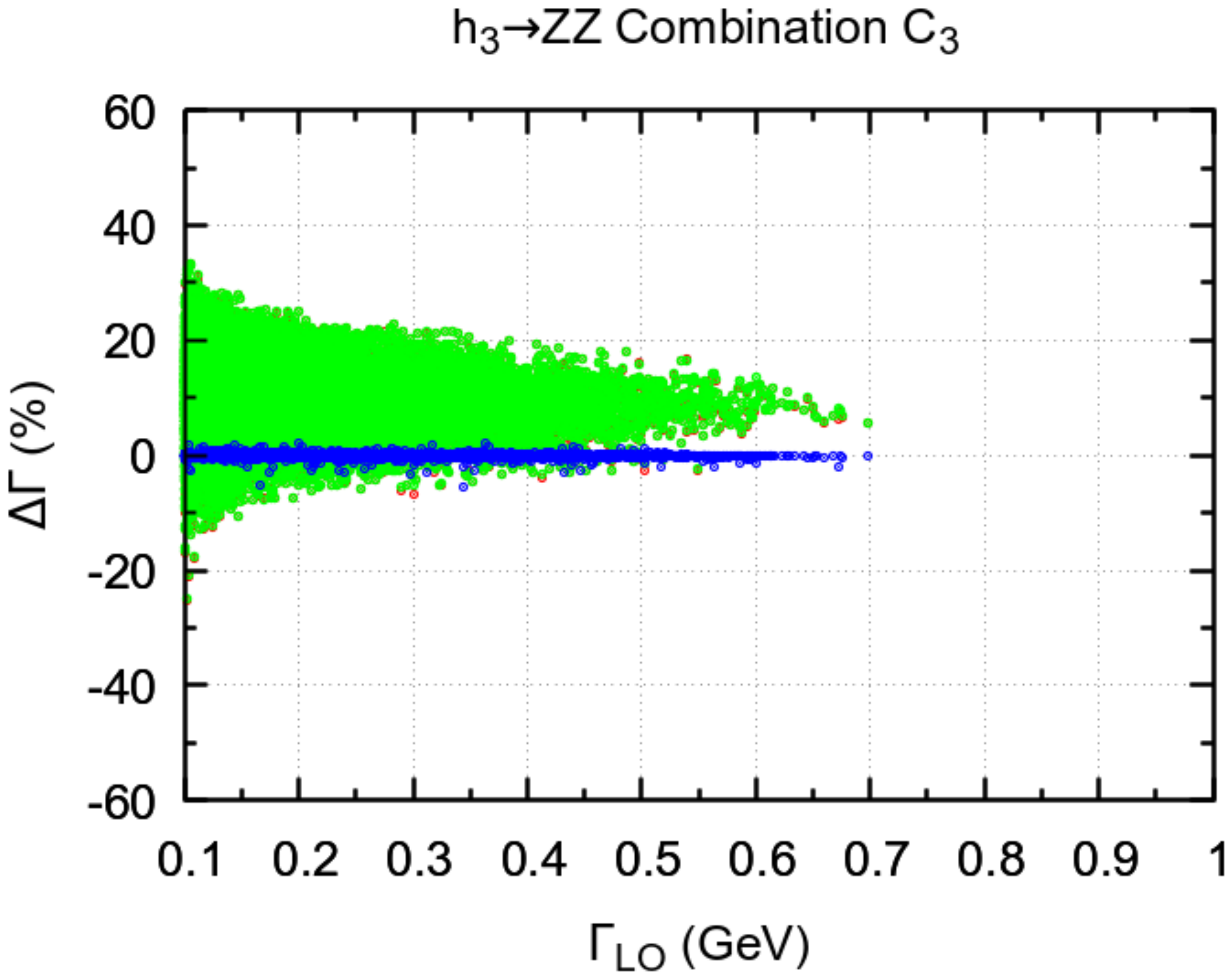}}
\caption{
$\Delta \Gamma_{h_3 \to ZZ}$ in percentage against $\Gamma^{\mathrm{LO}}_{h_3 \to ZZ}$, for the combinations $C_1$ (top), $C_2$ (bottom left) and $C_3$ (bottom right).
For $C_1$, we impose the cuts $\pi/2 -|\alpha_3| > 0.0005$ and $|\alpha_3|> 0$, whereas for $C_2$ and $C_3$ we impose the cuts $\pi/2 -|\alpha_3| > 0.005$ and $|\alpha_3|> 0$.
Only the interval $ 0.1 \, \, \textrm{GeV} < \Gamma^{\mathrm{LO}}_{h_3 \to ZZ} < 1 \, \, \textrm{GeV}$ is shown. 
In red, $\Delta \Gamma_{h_3 \to ZZ}$;
in green, $\Gamma^{\mathrm{mix}}_{h_3 \to ZZ} / \Gamma^{\mathrm{LO}}_{h_3 \to ZZ}$; in blue, $\Gamma^{m_3}_{h_3 \to ZZ} / \Gamma^{\mathrm{LO}}_{h_3 \to ZZ}$.
According to eq. \ref{eq:Gamma-def}, the red points are the sum of the green and the blue ones.}
\label{fig:h3ZZ}
\end{figure}

In figure \ref{fig:h3ZZ}, we show $\Delta \Gamma_{h_3 \to ZZ}$ in percentage against $\Gamma^{\mathrm{LO}}_{h_3 \to ZZ}$ for the combinations $C_1$, $C_2$ and $C_3$.
Whereas the red points represent the total correction, the green ones represent the contribution coming from $\Gamma^{\mathrm{mix}}_{h_3 \to ZZ}$ and the blue ones the contribution coming from $\Gamma^{m_3}_{h_3 \to ZZ}$.
We avoid the scenarios $S_1$ and $S_2$ by imposing the cuts described in the caption.

It is clear that, in all the plots, smaller values of $\Gamma^{\mathrm{LO}}_{h_3 \to ZZ}$ allow larger values of $\Delta \Gamma_{h_3 \to ZZ}$; the explanation is simply that, as the denominator of \ref{eq:Gamma-def} becomes smaller, the numerator does not necessarily mimic that reduction. Notice that whereas small values of $\Delta \Gamma_{h_3 \to ZZ}$ imply that a perturbative description of the theory is possible, large values would in principle require the calculation of the following order in perturbation theory, so as to ascertain the feasibility of such description.

As expected, the contribution arising from $\Gamma^{\mathrm{mix}}_{h_3 \to ZZ}$ in all the plots is similar to the equivalent one in the decay $h_2 \to ZZ$, described in ref. \cite{Fontes:2021iue}. Where the two processes differ is in the component $\Gamma^{m_3}_{j}$, which was absent in $h_2 \to ZZ$ but is present in $h_3 \to ZZ$, as discussed above. However, since we are requiring the points to be away from the troublesome scenarios $S_1$ and $S_2$, $\Gamma^{m_3}_{h_3 \to ZZ}$ generally takes small values. Accordingly, the total value $\Delta \Gamma_{h_3 \to ZZ}$ is essentially given by the contribution from $\Gamma^{\mathrm{mix}}_{h_3 \to ZZ}$ (which explains the fact that the red points are almost entirely hidden under the green ones).

\subsubsection{$h_3 \to h_1 Z$}
\label{sec:h3h1Z}

\begin{figure}[h!]
\centering
\subfloat{\includegraphics[width=0.55\linewidth]{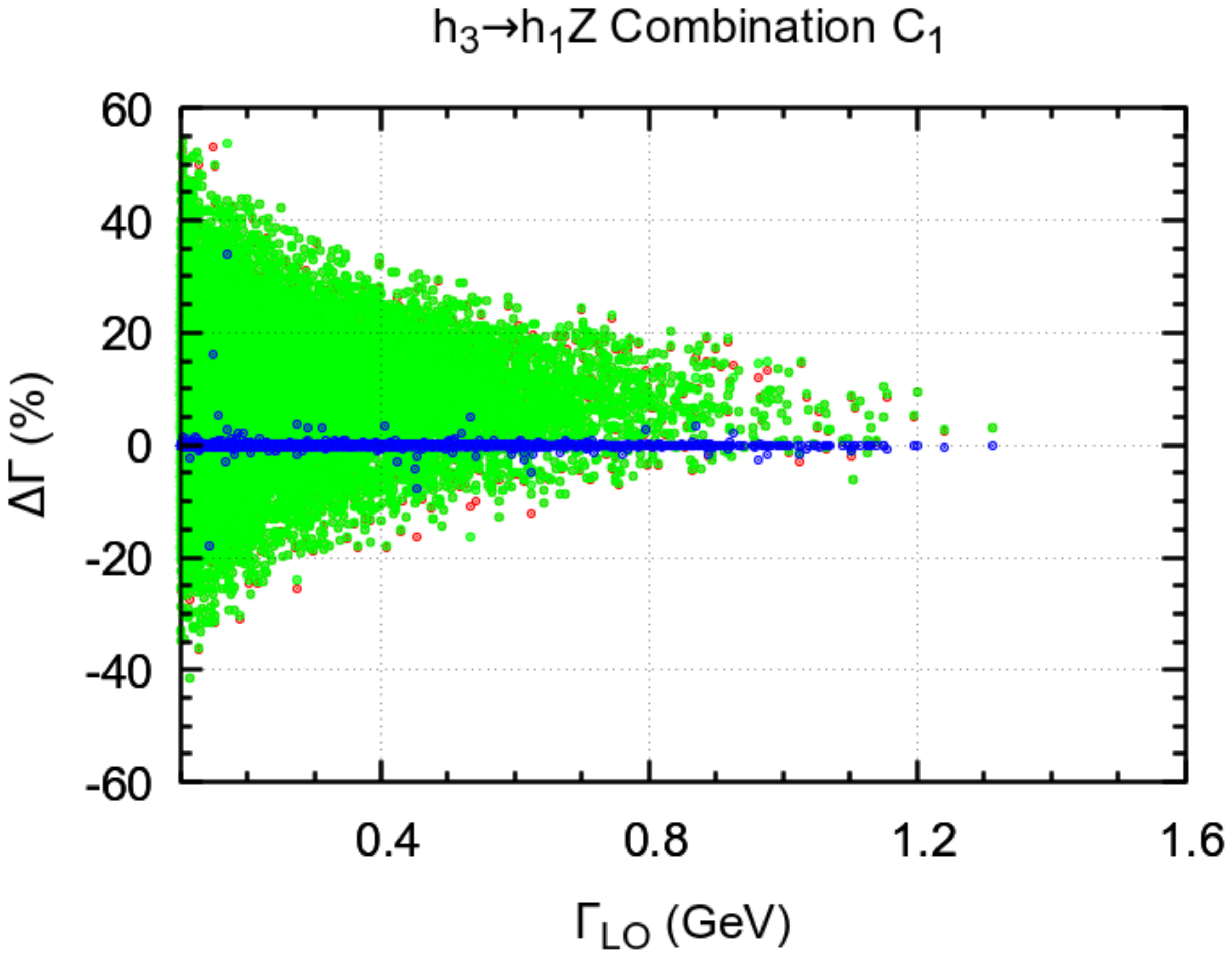}}\\
\subfloat{\includegraphics[width=0.49\linewidth]{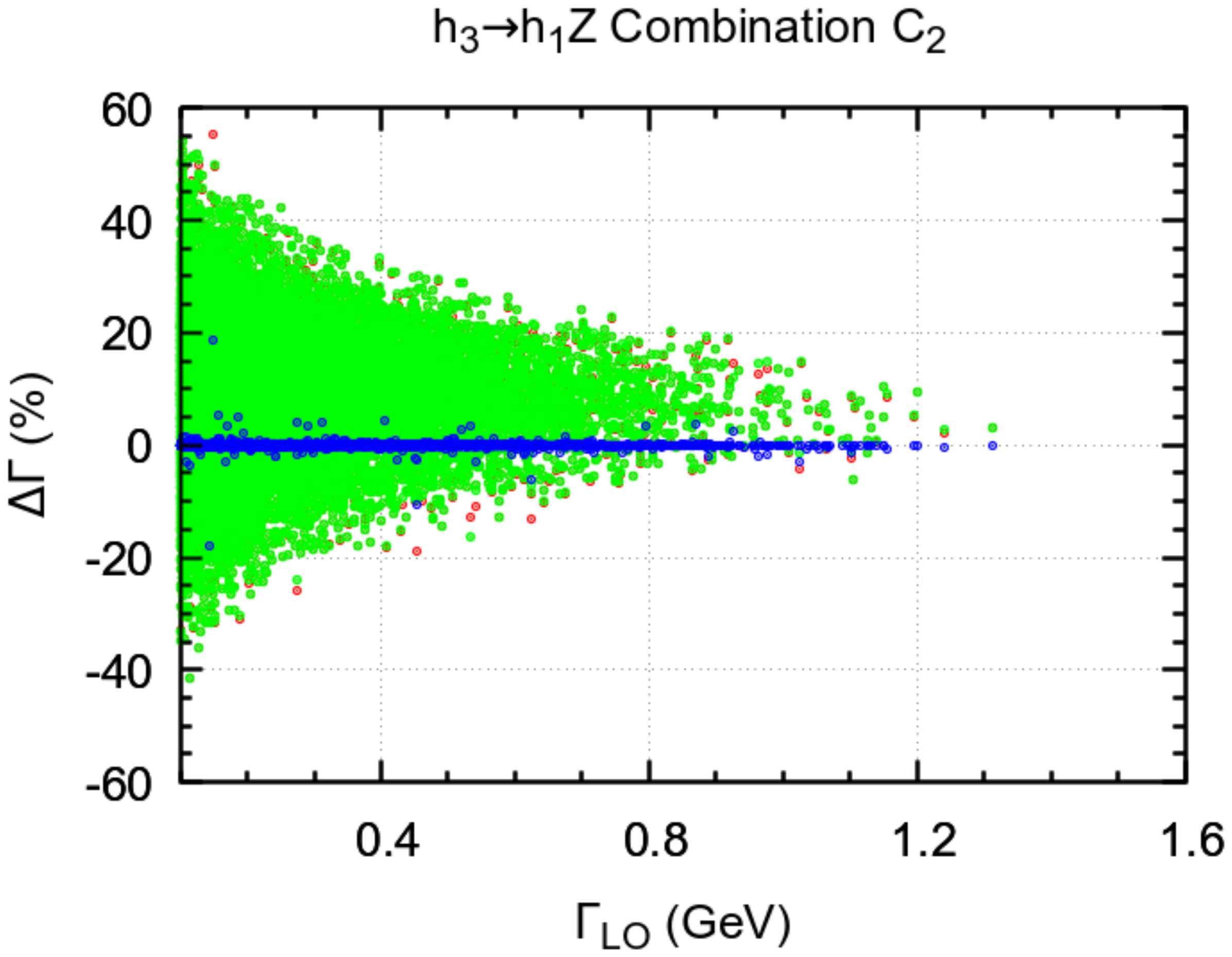}}\ \
\subfloat{\includegraphics[width=0.49\linewidth]{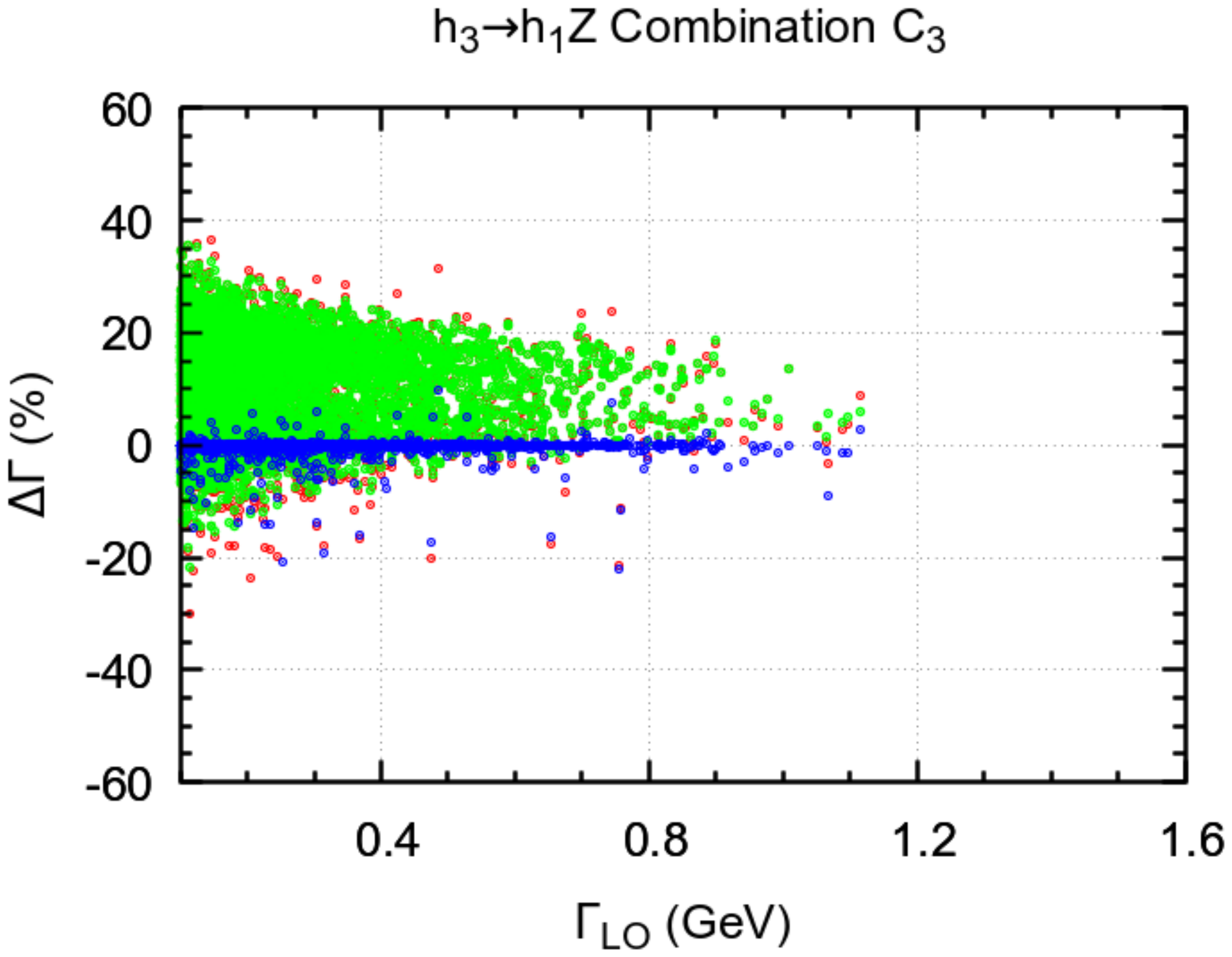}}
\caption{
$\Delta \Gamma_{h_3 \to h_1Z}$ in percentage against $\Gamma^{\mathrm{LO}}_{h_3 \to h_1Z}$, for the combinations $C_1$ (top), $C_2$ (bottom left) and $C_3$ (bottom right).
The cuts leading to the blue points in figure \ref{fig:Delta-m3-corrected} were applied.
Only the interval $ 0.1 \, \, \textrm{GeV} < \Gamma^{\mathrm{LO}}_{h_3 \to h_1Z} < 1.6 \, \, \textrm{GeV}$ is shown. 
In red, $\Delta \Gamma_{h_3 \to h_1Z}$;
in green, $\Gamma^{\mathrm{mix}}_{h_3 \to h_1Z} / \Gamma^{\mathrm{LO}}_{h_3 \to h_1Z}$; in blue, $\Gamma^{m_3}_{h_3 \to h_1Z} / \Gamma^{\mathrm{LO}}_{h_3 \to h_1Z}$.
According to eq. \ref{eq:Gamma-def}, the red points are the sum of the green and the blue ones.
}
\label{fig:h3h1Z}
\end{figure}

The results for $h_3 \to h_1 Z$ are similar to those of $h_3 \to Z Z$, as can be seen in figure \ref{fig:h3h1Z}.
For each combination, we apply the cuts associated to the blue points of the corresponding panel in figure \ref{fig:Delta-m3-corrected}. 
As before, $\Gamma^{m_3}_{h_3 \to ZZ}$ generally takes small values.

\subsubsection{$h_3 \to h_2 Z$}
\label{section:h3h2Z}

%

\begin{figure}[h!]
\centering
\subfloat{\includegraphics[width=0.55\linewidth]{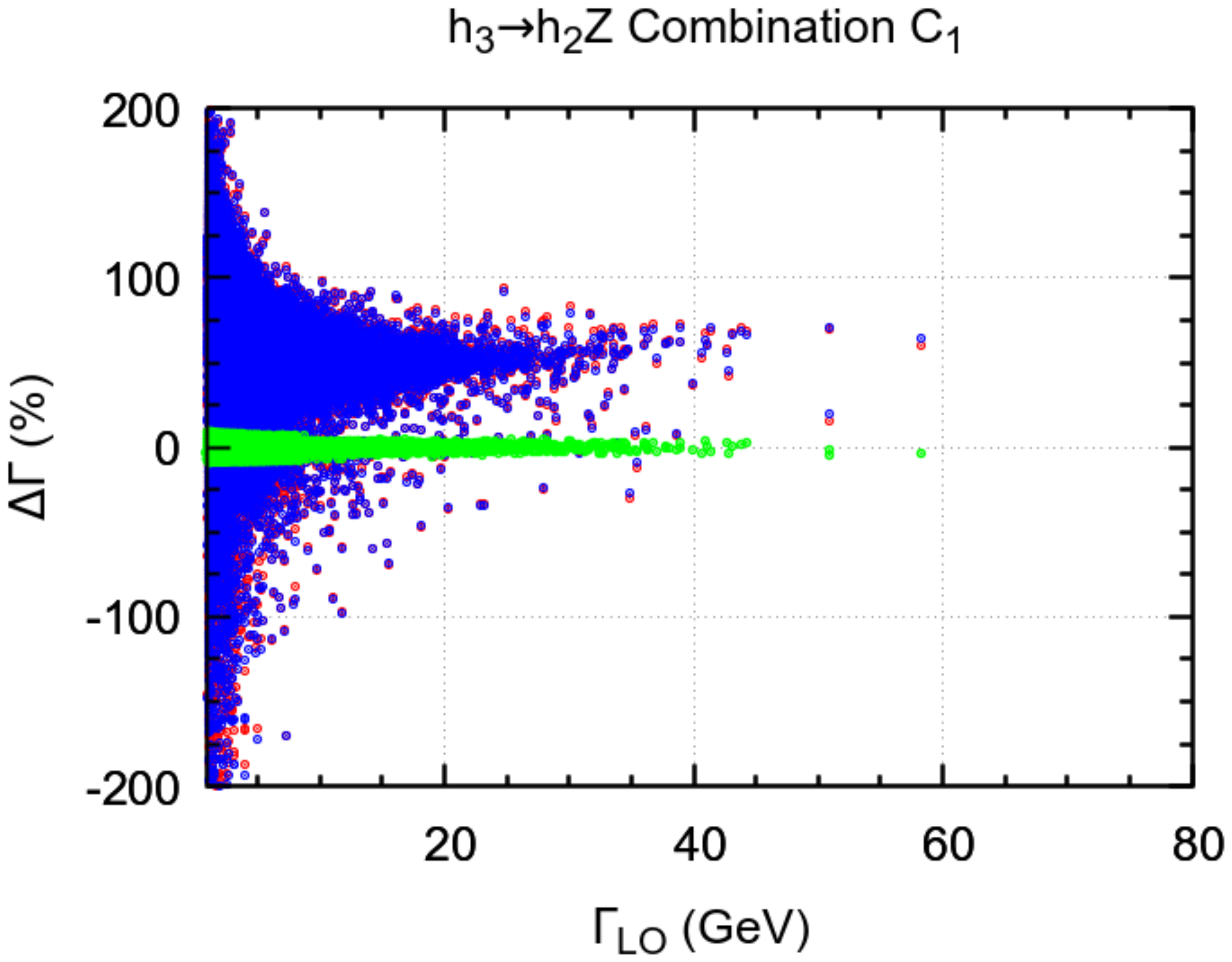}}\\
\ \
\subfloat{\includegraphics[width=0.48\linewidth]{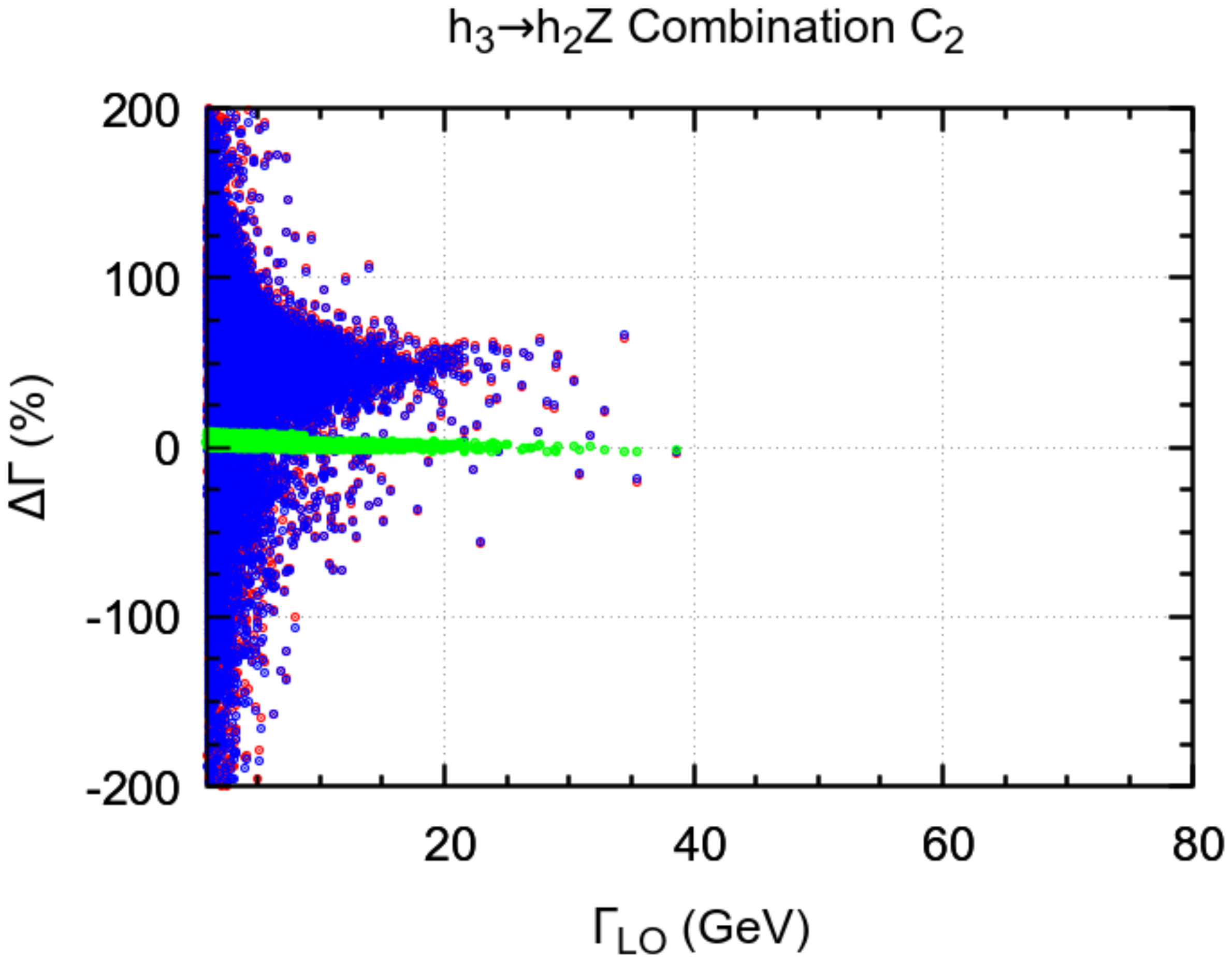}} \hspace{1mm}
\subfloat{\includegraphics[width=0.48\linewidth]{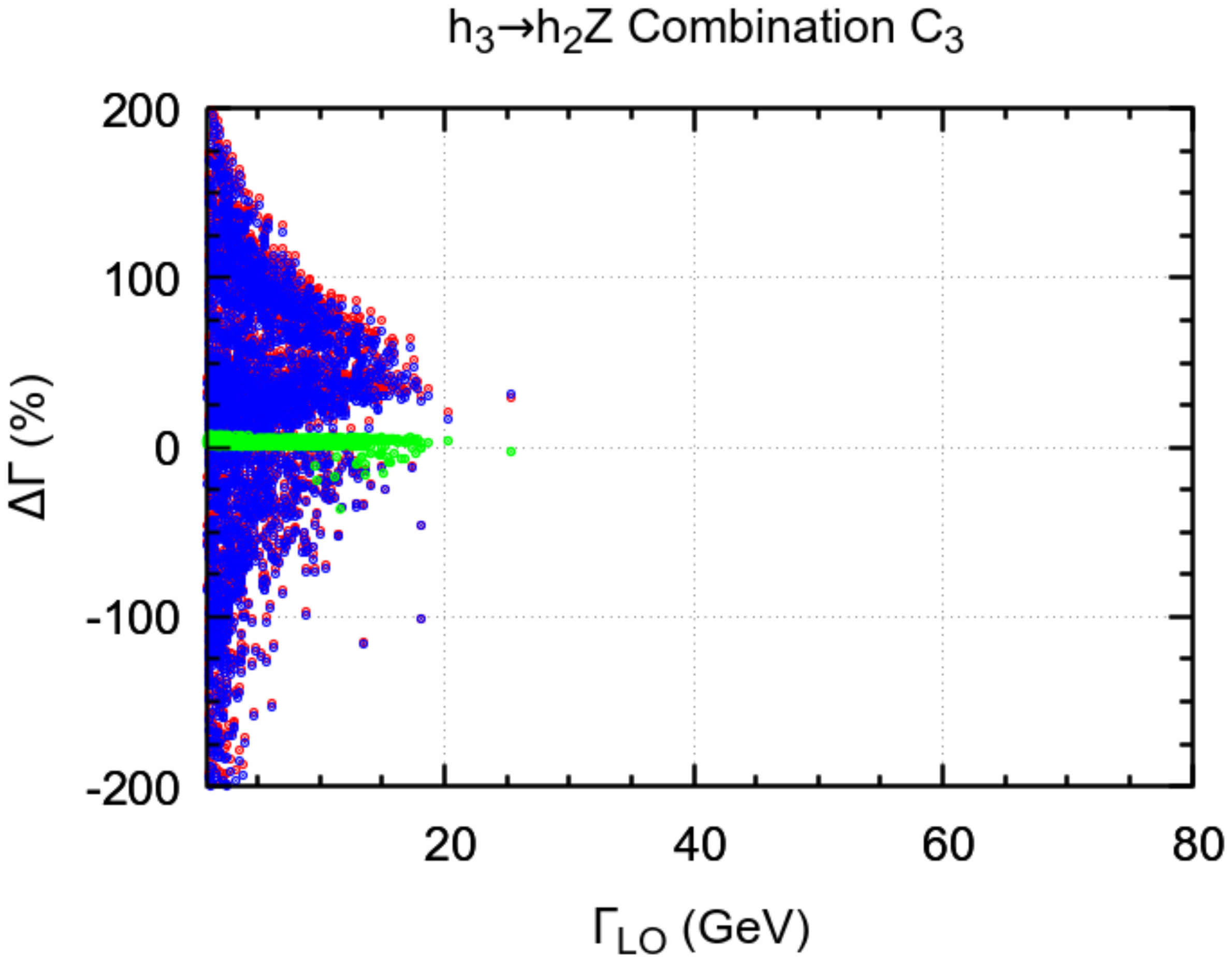}}
\caption{
$\Delta \Gamma_{h_3 \to h_2Z}$ in percentage against $\Gamma^{\mathrm{LO}}_{h_3 \to h_2Z}$, for the combinations $C_1$ (top), $C_2$ (bottom left) and $C_3$ (bottom right).
The cuts leading to the green points in figure \ref{fig:Delta-m3-corrected} were applied.
Only the interval 1 GeV $< \Gamma^{\mathrm{LO}}_{h_3 \to h_2Z} <$ 80 GeV is shown. 
In red, $\Delta \Gamma_{h_3 \to h_2Z}$;
in green, $\Gamma^{\mathrm{mix}}_{h_3 \to h_2Z} / \Gamma^{\mathrm{LO}}_{h_3 \to h_2Z}$; in blue, $\Gamma^{m_3}_{h_3 \to h_2Z} / \Gamma^{\mathrm{LO}}_{h_3 \to h_2Z}$.
According to eq. \ref{eq:Gamma-def}, the red points are the sum of the green and the blue ones.
}
\label{fig:h3h2Z}
\end{figure}

As already suggested, the decay $h_3 \to h_2 Z$ is significantly different from the other decays. This can be seen in figure \ref{fig:h3h2Z}, where we are using the same color code that was used in figures \ref{fig:h3ZZ} and \ref{fig:h3h1Z}. For each combination, we apply the cuts associated to the green points of the corresponding panel in figure \ref{fig:Delta-m3-corrected}. 
Whereas in figures \ref{fig:h3ZZ} and \ref{fig:h3h1Z}  the contribution from $\Gamma^{m_3}_{j}$ was small
(so that the total relative correction $\Delta \Gamma_j$ was essentially given by the component $\Gamma^{\mathrm{mix}}_{j}$), in figure \ref{fig:h3h2Z} we have precisely the opposite: $\Gamma^{\mathrm{mix}}_{h_3 \to h_2Z}$ is relatively small, whilst $\Gamma^{m_3}_{h_3 \to h_2Z}$ is by far the main contribution to $\Delta \Gamma_{h_3 \to h_2Z}$. Moreover, whereas $\Delta \Gamma_j$ in the previous figures barely reached $40 \%$ (and only for vary small values of $\Gamma^{\mathrm{LO}}_j$), $\Delta \Gamma_{h_3 \to h_2Z}$ can take values larger than that for the whole range of $\Gamma^{\mathrm{LO}}_{h_3 \to h_2Z}$.

\begin{figure}[h!]
\centering
\subfloat{\includegraphics[width=0.49\linewidth]{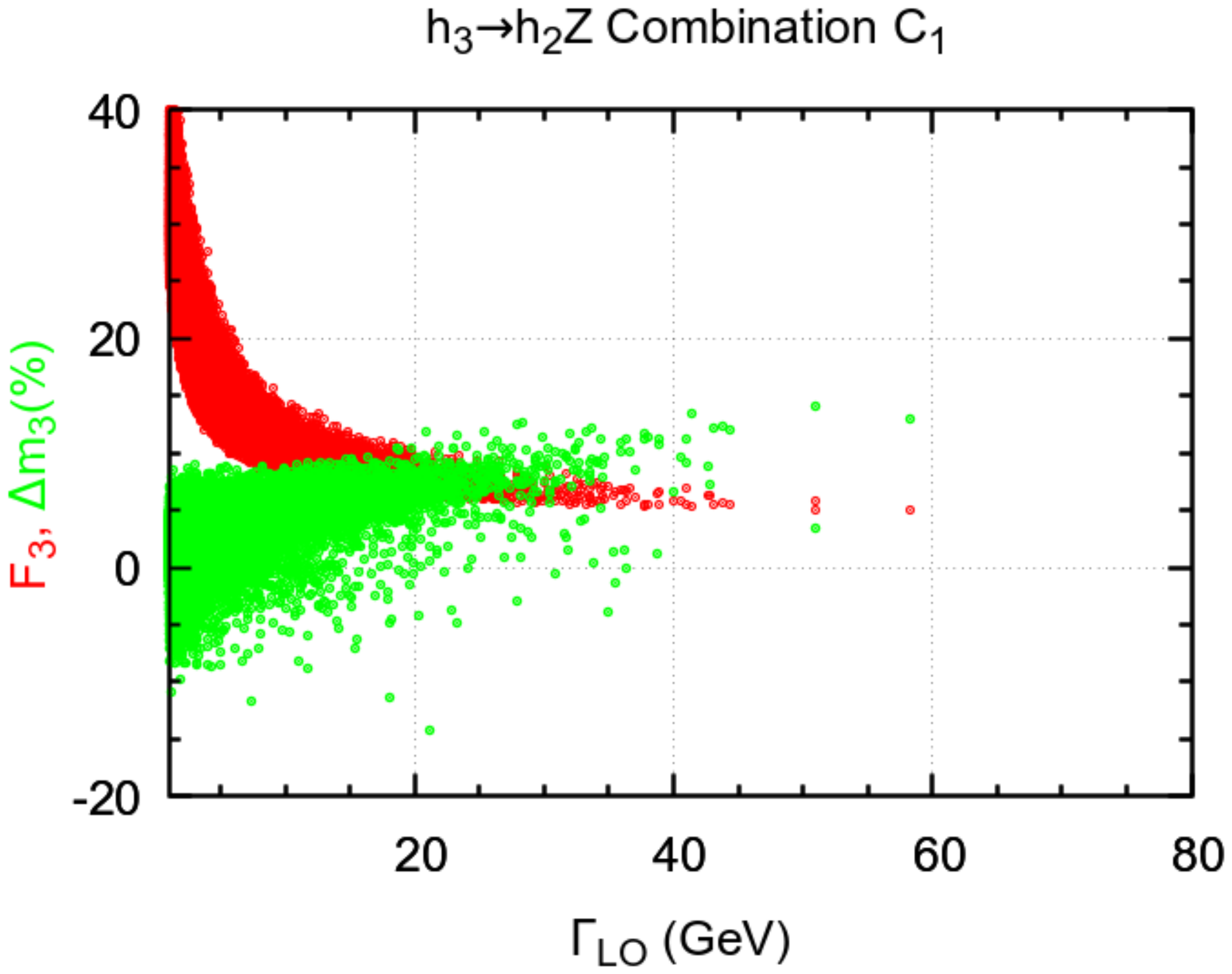}}
\ \
\subfloat{\includegraphics[width=0.49\linewidth]{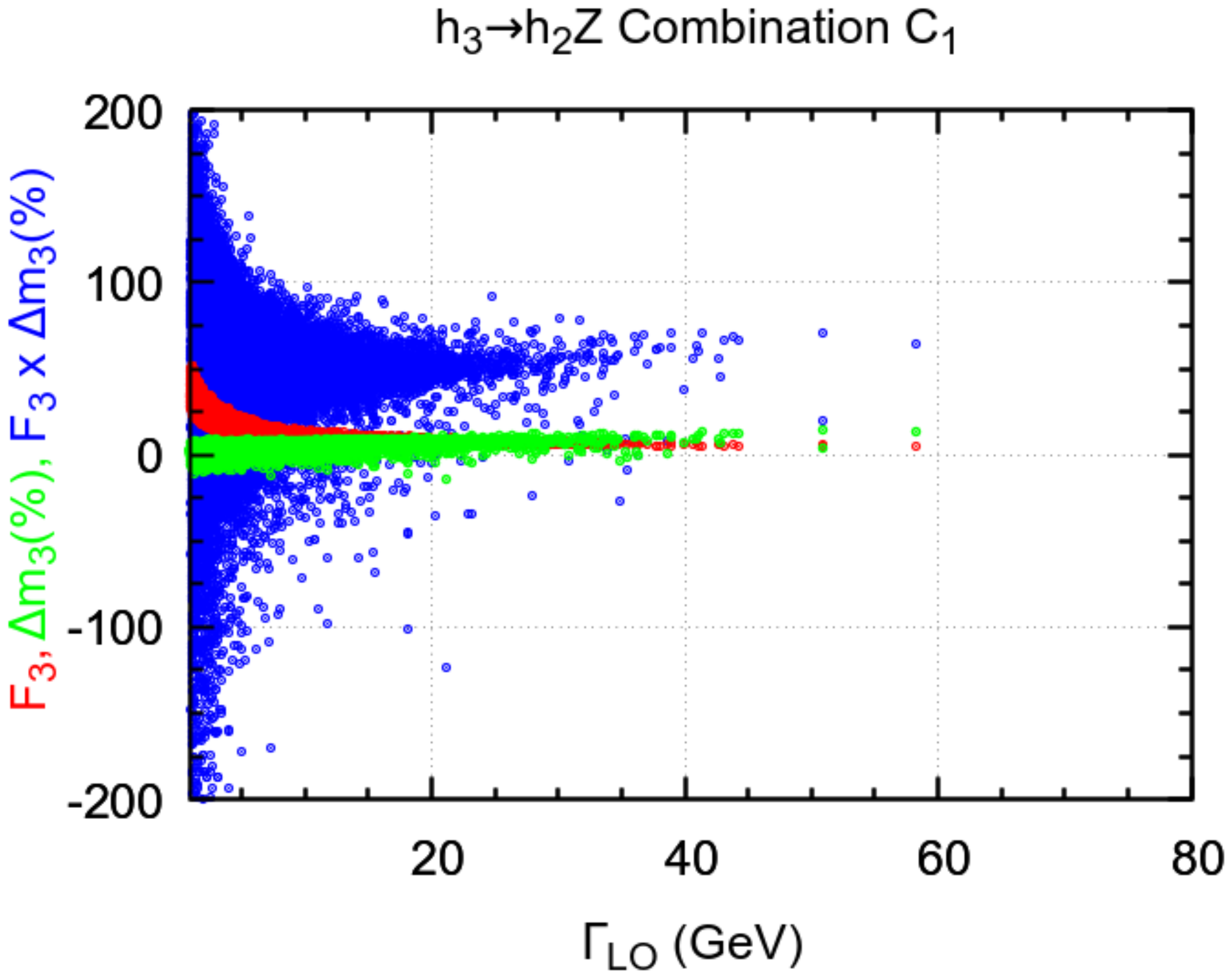}}
\caption{
Left: $F_{h_3 \to h_2 Z}$ (red) and $\Delta m_3$ (green) against $\Gamma^{\mathrm{LO}}_{h_3 \to h_2Z}$ for the combination $C_1$.
Right: the same, but also with the product (blue) of the green and red points. 
}
\label{fig:F-against-Dm}
\end{figure}

Since it is evident that those very large values are a consequence of very large values of $\Gamma^{m_3}_{h_3 \to h_2 Z}$, we now investigate this contribution. We can generically write $\Gamma^{m_3}_{j}$ as:
\be
\Gamma^{m_3}_{j}
=
F_{j} \, \, \Delta m_3 \, \, \Gamma^{\mathrm{LO}}_{j}.
\label{eq:rewritten}
\ee
Here, $F_{j}$ is an LO (thus combination-independent) dimensionless factor, which depends only on the tree-level masses of the particles involved in the decay $j$.
Eq. \ref{eq:rewritten} implies that the contribution from $\Gamma^{m_3}_{j}$ to $\Delta \Gamma_j$ (blue points in figures \ref{fig:h3ZZ}, \ref{fig:h3h1Z} and \ref{fig:h3h2Z}) is simply given by the product $F_{j} \, \, \Delta m_3$; that is,
\be
\dfrac{\Gamma^{m_3}_{j}}{\Gamma^{\mathrm{LO}}_j}
=
F_{j} \, \, \Delta m_3.
\label{eq:Gamma-var}
\ee
This means that the impact of $\Gamma^{m_3}_{j}$ on $\Delta \Gamma_j$ is given by the NLO correction $\Delta m_3$ weighted by the LO factor $F_j$. As a result, even if $\Delta m_3$ does not take large values (and so behaves perturbatively), the impact of $\Gamma^{m_3}_{j}$ on $\Delta \Gamma_j$ may end up being large (and so non-perturbative) if $F_j$ turns out to be non-negligible.
%
%
Now, from eq. \ref{eq:Gm3-h3h2Z}, one can read:
\be
F_{h_3 \to h_2 Z} =
-3 \, \dfrac{m_2^4 - m_{3\mathrm{R}}^4 - 2 \, m_2^2 \, m_{\textrm{Z}}^2 + m_{\textrm{Z}}^4 }{m_{3\textrm{R}}^4 + m_2^4 + m_{\textrm{Z}}^4 - 2 m_{3\textrm{R}}^2 m_2^2 - 2 m_{3\textrm{R}}^2 m_{\textrm{Z}}^2 - 2 m_2^2 m_{\textrm{Z}}^2}.
\ee
On the left plot of figure \ref{fig:F-against-Dm},
we show $F_{h_3 \to h_2 Z}$ (in red) and $\Delta m_3$ (in green) against $\Gamma^{\mathrm{LO}}_{h_3 \to h_2Z}$ for the combination $C_1$.
We see that $F_{h_3 \to h_2 Z}$ always takes values larger than 5 for the whole range of $\Gamma^{\mathrm{LO}}_{h_3 \to h_2Z}$. 
Hence, the impact of $\Gamma^{m_3}_{h_3 \to h_2 Z}$ on $\Delta \Gamma_{h_3 \to h_2 Z}$ is at least 5 times the NLO correction $\Delta m_3$. This is clearly shown on the right plot, where the points in blue represent the product between the red and the green points (note that, according to eq. \ref{eq:Gamma-var}, those blue points are precisely the blue points on the top panel of figure \ref{fig:h3h2Z}). This explains the large values (\hspace{-1mm}$\gsim 40 \%$) for $\Delta \Gamma_{h_3 \to h_2 Z}$ observed in figure \ref{fig:h3h2Z}. As we just suggested, these non-perturbative results do not stem from non-perturbative corrections to the $h_3$ mass; rather, even though such corrections take moderate values (green points in the left panel of figure \ref{fig:F-against-Dm}), they end up being several times enhanced due to the peculiar LO factor $F_{h_3 \to h_2 Z}$.

\subsubsection{$h_3 \to h_2 h_1$}
\label{sec:h3h2h1}

\begin{figure}[h!]
\centering
\includegraphics[width=0.60\linewidth]{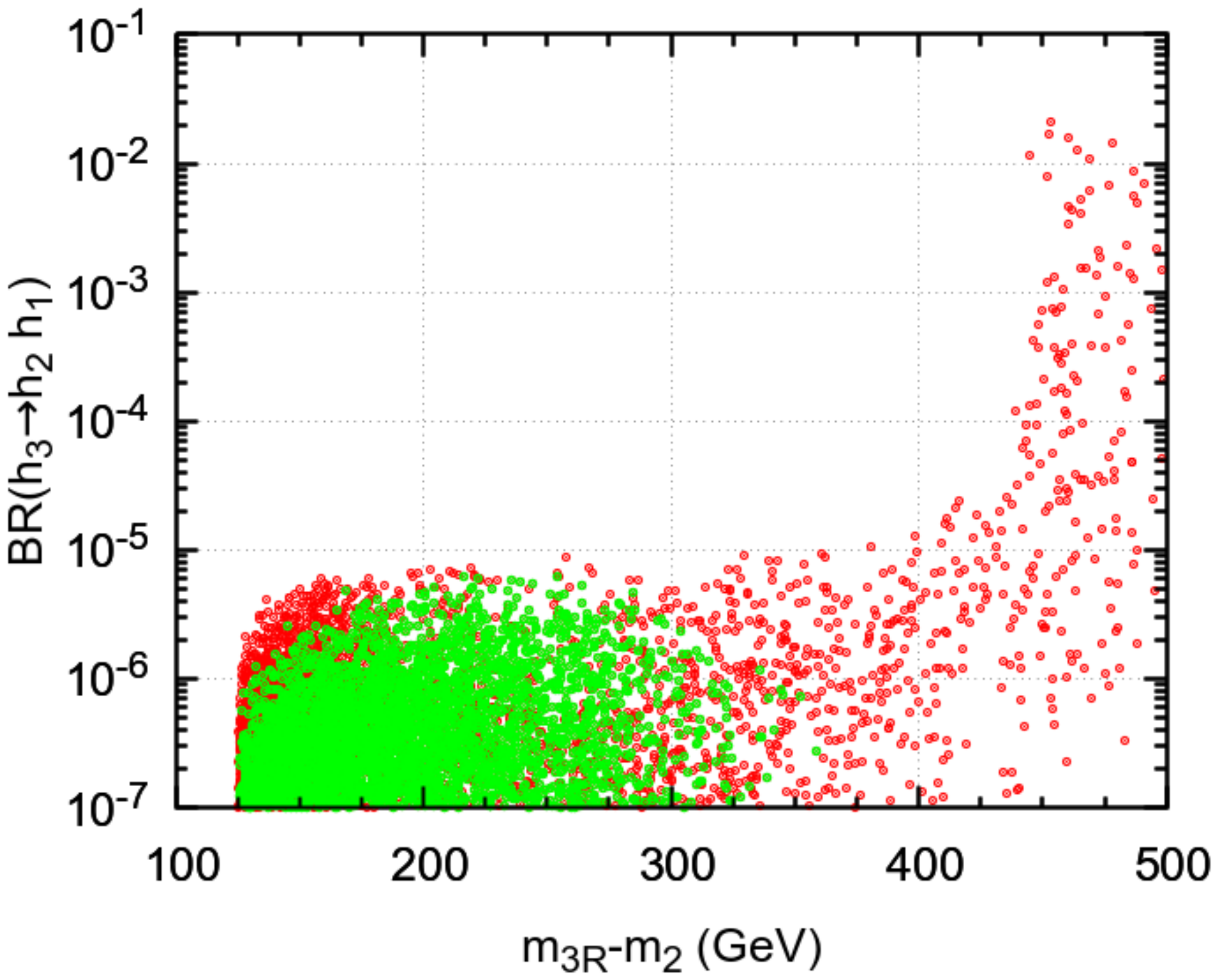}
\caption{Tree-level branching ratio of $h_3 \to h_2 h_1$ against the difference $m_{3\mathrm{R}} - m_2$. In red, points passing all constraints except \ts{HiggsBounds5}; in green, points passing all constraints.}
\label{fig:BR-h3h2h1}
\end{figure}

The decay $h_3 \to h_2 h_1$ can be especially interesting, because its discovery would constitute an undoubtable sign of CP violation in the scalar sector (in the specific context of a 2HDM) \cite{Fontes:2015xva,Low:2020iua}.%
\fn{This process does not necessarily imply CP violation in models other than the 2HDM \cite{Fontes:2015xva}.}
In ref. \cite{Low:2020iua}, it was claimed that, at least for a very fine-tuned region of the parameter space, the process can have a branching ratio of about $3\%$ at tree-level. We were not able to find such region. To be sure, we do find regions of theoretically valid points that lead to a tree-level branching ratio of the order of the percent level; only, they end up being ruled out by experimental results included in \ts{HiggsBounds5}.%
\fn{These include a variety of processes, but the most important ones are $p p \to (H^+) t \overline{b} \to (t \overline{b}) t \overline{b}$ \cite{ATLAS:2020jqj} and $p p \to h_3 \to V V$ \cite{ATLAS:2018sbw}.}
This behaviour can be seen in figure \ref{fig:BR-h3h2h1}, where we plot points with and without the constraints coming from \ts{HiggsBounds5}. The figure clearly shows that, after applying such constraints, the branching ratio does not even reach $0.001\%$.

One might wonder
whether the inclusion of one-loop effects may significantly change these results.
Although it is expected that the contributions from eqs. \ref{eq:h3-h2h1} do not lead to big corrections to the tree-level decay width (as they are expected to show a perturbative behaviour), the one-loop corrections to $m_{3\mathrm{R}}$ could in principle lead to a broadening of the region of parameter space where the decay $h_3 \to h_2 h_1$ is kinematically allowed, which could allow regions with an enhanced decay width. We found that this is not the case; in fact, even considering the corrected $h_3$ mass, we can still find no regions of the parameter space leading to a relevant decay width for $h_3 \to h_2 h_1$.

\section{Conclusions}
\label{sec:conclusions}

We discussed NLO corrections to the derived mass $m_3$ of the heaviest neutral scalar $h_3$ in the C2HDM---one of the most simple models beyond the SM with CP violation in the scalar sector---and investigated the impact of such corrections in NLO decays of $h_3$. 
Due to the presence of CP violation in the scalar sector of the C2HDM, the renormalization of the model requires the introduction of several non-physical parameters, so that one ends up having different possible combinations of independent counterterms for the same set of independent renormalized parameters.
Restricting ourselves to four combinations, we found that the corrections to $m_3$ can be extremely large in all of them. The origin of problem lies in the particular dependence of $m_3$ on the independent parameters. Such dependence, in fact, leads to very large values in certain limits; this happens in such manner that, in very fine-tuned regions of the parameter space, the LO mass ends up assuming moderate values, whereas the NLO corrections become very large. We showed that, outside the fine-tuned regions, the corrections acquire moderate 
values.

We then investigated four specific NLO decay widths of $h_3$. We stressed the existence of a contribution to those decay widths arising from the NLO corrections to the mass of $h_3$. We showed that such contribution is small in  $h_3 \to ZZ$ and $h_3 \to h_1 Z$, as long as the aforementioned fine-tuned regions are avoided. In the case of $h_3 \to h_2 Z$, however, there is a large superposition between those regions and the ones which kinematically allow the decay. As a consequence, the NLO corrections to the mass must become larger, although still perturbative; it turns out that a LO multiplicative factor enhances them, leading to NLO corrections to the decay width of the order of $50\%$.
Finally, we discussed the process $h_3 \to h_2 h_1$; we showed that, although its discovery would constitute an irrefutable sign of CP violation in the scalar sector of a 2HDM, the current experimental results hinder relevant values for its decay width.

It would be of interest to explore the scenario where the renormalization of the model is performed by taking all the masses of the physical particles as independent parameters. One would then need to ascertain the behaviour of the counterterms; in particular, it would be relevant to investigate if there are regions of the parameter space for which the NLO corrections take large values, just as in the scenarios described in this paper.
It would also be interesting to see how the renormalization program described here can affect the relation between CP-violating phases and CP-violating observables previously studied in ref. \cite{Fontes:2017zfn}.

\section*{Acknowledgments}

We thank
Sally Dawson,
Ansgar Denner,
Florian Domingo,
Sven Heinemeyer,
Sebastian Paßehr,
João P. Silva
and 
Xiaoping Wang
for discussions.
DF was supported by the United States Department of Energy under Grant Contract DE-SC0012704, as well as by the Portuguese \textit{Funda\c{c}\~ao para a Ci\^encia e Tecnologia} (FCT) under the project SFRH/BD/135698 /2018. JCR was supported by FCT through the projects CFTP-FCT Unit 777 (UIDB/00777 /2020
and UIDP/00777/2020) and PTDC/FIS-PAR/29436/2017,
which are partially funded through POCTI (FEDER), COMPETE, QREN and EU.

\vspace*{1cm}
\bibliographystyle{JHEP}
\bibliography{Inputs/MyReferences}

\end{document}